\pdfoutput=1
\documentclass[iop]{emulateapj}
\usepackage{apjfonts}
\usepackage{amssymb}
\usepackage{amsmath}
\usepackage{ctable}

\newcommand{\be}{\begin{equation}}
\newcommand{\ee}{\end{equation}}

\newcommand{\rhogas}{\rho_{\rm g}}
\newcommand{\grain}{_{\rm p}}
\newcommand{\rhointernal}{\bar{\rho}_{\rm solid}}

\newcommand{\cs}{c_{s}}
\newcommand{\Rgrain}{a}
\newcommand{\tstop}{t_{\rm s}}
\newcommand{\taustop}{\tau_{\rm s}}

\newcommand{\eddy}{_{e}}

\newcommand{\Teddy}{t\eddy}

\newcommand{\vspacerpostplot}{\vspace{0.1cm}}

\newcommand\plotonesize[2]
 {\centering \leavevmode \includegraphics[width={#2\columnwidth}]{#1}}

\shorttitle{Dust-to-Gas Fluctuations \&\ Metal Stars}
\shortauthors{Hopkins \&\ Conroy}
\slugcomment{Submitted to ApJ, December, 2015}
\begin{document}
\title{\vspace{-1.0cm}Are the Formation and Abundances of Metal-Poor Stars the Result of Dust Dynamics?}
\author{Philip F.\ Hopkins\altaffilmark{1} \&\ Charlie Conroy\altaffilmark{2}}
\altaffiltext{1}{TAPIR, Mailcode 350-17, California Institute of Technology, Pasadena, CA 91125, USA; E-mail:phopkins@caltech.edu}
\altaffiltext{2}{Harvard-Smithsonian Center for Astrophysics, 60 Garden St., Cambridge, MA 02138, USA}

\label{firstpage}

\begin{abstract}

Large dust grains can fluctuate dramatically in their local density, relative to the gas, in neutral, turbulent disks. Small, high-redshift galaxies (before reionization) represent ideal environments for this process. We show via simple arguments and simulations that order-of-magnitude fluctuations are expected in local abundances of large grains ($>100\,$\AA) under these conditions. This can have important consequences for star formation and stellar metal abundances in extremely metal-poor stars. Low-mass stars could form in dust-enhanced regions almost immediately after some dust forms, even if the galaxy-average metallicity is too low for fragmentation to occur. We argue that the metal abundances of these ``promoted'' stars may contain interesting signatures, as the {\small CNO} abundances (concentrated in large carbonaceous grains and ices) and {\small Mg} and {\small Si} (in large silicate grains) can be enhanced and/or fluctuate almost independently. Remarkably, the otherwise puzzling abundance patterns of some metal-poor stars can be well-fit by standard, IMF-averaged core-collapse SNe yields, if we allow for fluctuating local dust-to-gas ratios. We also show that the observed log-normal-like distribution of enhancements in these species agrees with our simulations. Moreover, we confirm {\small Mg} and {\small Si} are correlated in these stars, with abundance ratios similar to those in local silicate grains. Meanwhile {\small [Mg/Ca]}, predicted to be nearly invariant from pure SNe yields, shows very large enhancements and variations up to factors $\gtrsim100$ as expected in the dust-promoted model, preferentially in the {\small [C/Fe]}-enhanced metal-poor stars. Together, this suggests that (1) dust exists in second-generation star formation, (2) local dust-to-gas ratio fluctuations occur in proto-galaxies and can be important for star formation, and (3) the light element abundances of these stars may be affected by the local chemistry of dust where they formed, rather than directly tracing nucleosynthesis from earlier populations. 

\end{abstract}


\section{Introduction}
\label{sec:intro}

Extremely metal-poor stars represent a laboratory for studying the conditions in the early Universe. They are sensitive probes of stellar evolution and supernovae (SNe) explosion models, nucleosynthesis and the origin of elements heavier than {\small H} and {\small He}, the enrichment and early formation history of galaxies, and the nature and origins of the first stars. Their metal abundance patterns present many outstanding challenges and unsolved problems; understanding the origin of these stars and their stellar abundances is critical to all of the open questions above. 

For example, most of the observed extremely-metal poor ({\small [Fe/H]}$<-3.0$) population appears to be dramatically enhanced in light ({\small CNO}) elements ($> 40\%$ and $> 80\%$ of stars below {\small [Fe/H]}$<-3.0$ and $-4.0$; \citealt{placco:2014.cemp.classes}), including the various sub-classes of carbon-enhanced metal-poor stars (CEMPS) with {\small [C/Fe]}\,$\sim0-4$ \citep{lee.2013:metal.poor.star.cemp.frequency}. A plausible explanation for the carbon enhancements is essentially a selection effect: low-mass star formation requires efficient cooling (to allow collapse and fragmentation); clouds with too low a total metal mass {\small [Z/H]}\,$\ll -3$ would either form stars very inefficiently, or only form massive (short-lived) stars \citep{schneider:2003.low.mass.cooling.channels,omukai:2005.fragmetation.at.low.metallicity,chiaki:2014.critical.dust.abundance.for.cooling,ji:2014.si.dust.cooling.threshold.for.early.stars,ji:2014.silicate.dust.cooling.for.metal.poor.star.criterion.and.tests}, so stars selected with very low {\small [Fe/H]} ``should,'' in this argument, have high abundances in the {\small CNO} species constituting most of the metal mass. \footnote{In this paper, we will generally use ``CEMP'' as a shorthand for  {\small CNO}-enhanced, extremely metal poor stars, including NEMP and OEMP stars, which are believed to form early in the Universe and with metal abundances reflecting their formation conditions rather than stellar evolution or pollution by a binary companion. This means we focus on the CEMP-no population, dominant at {\small [Fe/H]}\,$<-3$, rather than the CEMP-r, CEMP-s, and CEMP-r/s populations \citep[for extended discussion, see][]{carollo.2014:cemp.bimodality,hansen.2015:cemp.subclassifications.review,maeder.2015:cemp.classifications,bonifacio.2015:cemp.bimodality}.} But this only explains why we might preferentially see CEMPS today, not how the enhancement (e.g.\ high {\small [C/Fe]}) was produced in the first place. And it does not {\em necessarily} tell us anything about the galaxies or environments in which these stars formed, only about the local cloud which collapsed to form the star. Other light elements such as {\small Si} and {\small Mg} also display unusual enhancements and correlations which are not well-understood \citep[see e.g.][]{aoki:2002.cemp.mg.si.abundances}; in ultra-faint dwarfs these also have been observed to vary and usually appear in excess together \citep{aoki:2002.cemp.mg.si.abundances,norris:2010.seg.1.cemp} but in limited cases may also vary independently \citep{norris:2010.bootes.cemp,frebel:2014.seg.1.cemp}. 

The simplest explanation for these metal abundances, that they reflect the yields of normal core-collapse SNe (averaged over the stellar initial mass function [IMF]), fails to predict anything like the observed stellar abundances of the extremely metal-poor stars ({\small [Fe/H]}$<-3$). Of course, at these low metallicities, the number of progenitor SNe enriching the ISM may be small, so models typically allow for metal-poor or metal-free progenitor stars, and an arbitrary mix of progenitor stellar masses (i.e.\ assuming the abundances might come from just one, or at most a few SNe, with individual explosion and progenitor parameters fitted to the observations). However even with these degrees of freedom, the models still often fail to explain the abundances of certain individual species at the order-of-magnitude level \citep[see e.g.][]{nomoto2006:sne.yields,heger.woosley:2010.metal.free.star.nucleosynthesis,lee.2014:cemp.production.normal.imf.plus.agb.stars.works.metal.rich,placco.2015:ump.fits.to.cno.elements} -- although they undoubtedly explain many of the observed abundance ratios. For the lowest-metallicity stars observed ({\small [Fe/H]}$<-4$), and in particular for the CEMP stars, these remaining discrepancies have led to more ``exotic'' models with a number of free parameters, invoking a mix of normal/faint SNe and hypernovae (with variable explosion energies $\sim 10^{51}-10^{54}\,{\rm erg}$), jets and prior ``failed explosions'' and fallback episodes, rotation and adjustable mixing layers allowing for a tunable stellar abundance profile in the progenitor stars, and pollution of the stars via companions \citep[e.g.][]{tominaga:popIII.nucleosynthesis,ishigaki.2014:faint.popIII.sne.cemps,takahashi.2014:metal.poor.explosion.yields,abate.2015:cemps.binary.pollution}. These additions can improve the agreement with observations; however, there is still no consistent theoretical scenario which simultaneously explains most of the observed stars, and even the best-fit models for many individual stars can still have order-of-magnitude discrepancies with certain ``outlier'' elements \citep[see][and references therein]{tominaga.2014:metal.poor.star.compilation,frebel.2015:another.metal.poor.star,placco.2015:ump.fits.to.cno.elements}.


An alternative explanation, therefore, which has thus far not been much discussed \citep[although see][]{gimore.2013:bimodal.metal.poor.formation.channels,maeder.2015:cemp.classifications}, is that the metallicities (and even abundance ratios) of regions which successfully form low-mass stars at extremely low metallicities do {\em not} necessarily reflect the direct SNe yields. This can happen via many distinct physical mechanisms. One mechanism which is particularly appealing, and already known to occur under the right conditions, is the segregation of dust and gas by aerodynamic drag. In short, in primarily neutral, dense gas, massive dust grains (which, at least at low redshift, contain a large fraction of all the ISM metal mass) behave as aerodynamic particles (i.e.\ they are coupled to the gas via drag from collisions with atoms or molecules). As such, they do not move perfectly with gas, and under certain circumstances they can clump or disperse independent of the gas, generating large fluctuations in the dust-to-gas ratio. By extension, this leads to fluctuations in the abundance of different species (those concentrated in grains of the correct size) from one region to another. 

In proto-planetary disks, this phenomenon is well-studied and is believed to be critical for planetesimal formation. When stirred by turbulence, ``trapped'' in long-lived vortices or pressure extrema (vortex or pressure traps), or excited by motions generated by the dust-gas interaction itself (the ``streaming'' instability), the local number density of dust grains can fluctuate by several orders of magnitude relative to gas \citep[see e.g.][]{bracco:1999.keplerian.largescale.grain.density.sims,cuzzi:2001.grain.concentration.chondrules,youdin.goodman:2005.streaming.instability.derivation,johansen:2007.streaming.instab.sims,bai:2010.grain.streaming.vs.diskparams,hopkins:2014.pebble.pile.formation}. The same phenomenon of ``preferential concentration'' is well-known in both laboratory experiments and simulations of terrestrial turbulence \citep{squires:1991.grain.concentration.experiments,rouson:2001.grain.concentration.experiment,monchaux:2010.grain.concentration.experiments.voronoi,monchaux:2012.grain.concentration.experiment.review,pan:2011.grain.clustering.midstokes.sims}. Because the dynamics are scale-free, the same phenomena governing pebbles and boulders in a proto-planetary disk should apply to sub-micron size dust in a giant molecular cloud \citep{padoan:dust.fluct.taurus.vs.sims,yoshimoto:2007.grain.clustering.selfsimilar.inertial.range,bec:2009.caustics.intermittency.key.to.largegrain.clustering,olla:2010.grain.preferential.concentration.randomfield.notes,hopkins:2013.grain.clustering}. And indeed, dust-to-gas ratio fluctuations on scales $\sim 0.003-10\,$pc have been measured in many nearby molecular clouds and some galaxy nuclei \citep[e.g.\ Orion, Taurus, and many more, see][]{thoraval:1997.sub.0pt04pc.no.cloud.extinction.fluct.but.are.on.larger.scales,thoraval:1999.small.scale.dust.to.gas.density.fluctuations,abergel:2002.size.segregation.effects.seen.in.orion.small.dust.abundances,miville-deschenes:2002.large.fluct.in.small.grain.abundances,padoan:dust.fluct.taurus.vs.sims,flagey:2009.taurus.large.small.to.large.dust.abundance.variations,pineda:2010.taurus.large.extinction.variations,nyland:2013.radio.core.ngc1266}, with a dependence on dust grain size in good agreement with the predictions of turbulent concentration \citep{padoan:dust.fluct.taurus.vs.sims,hopkins:totally.metal.stars}. As a result, \citet{hopkins:totally.metal.stars} speculated that these might lead to local variations in the metallicity of star-forming regions and metal abundances of the resulting stars. However, they concluded that, in the {\em local} Universe, the $\sim1\sigma$ scatter in metal abundances owing to these effects would be $<0.1\,$dex in most cases; order-of-magnitude effects would be extremely rare. 

In this paper, we consider instead how such fluctuations in the local dust-to-gas ratio might occur in high-redshift, predominantly neutral galaxies. Using both simulations and analytic arguments, we show that the effects of turbulent concentration may be much more dramatic in these proto-galaxies, leading (potentially) to large fluctuations in the local metallicity that can generate locally dust-rich and metal-rich regions which will preferentially cool, fragment, and form low-mass stars (``promoted'' star formation). Using a simple model for dust, we examine how this might alter the interpretation of the observed metal abundances of the stars. Remarkably, we show that some of the light-element ({\small CNO}, {\small Mg}, and {\small Si}) enhancements and correlations which are most difficult to reproduce in SNe nucleosynthesis models result naturally from the assumption that stars formed in unusually dust-rich regions.

\section{The Scenario}
\label{sec:model}

Before going into details, we briefly sketch the scenario we will explore. In a high-redshift mini-halo, one or more Pop-III SNe explode, providing a seed amount of dust ($-7\lesssim$\,{\small [Z/H]}\,$\lesssim -3$). That initial dust-to-gas ratio can then, under the right conditions, be enhanced by one or more orders of magnitude. The enhanced regions will then have local metallicities high enough for ``standard'' cooling, fragmentation and low-mass star formation to occur at high densities. Any low-mass (long-lived) stars that form this way will be imprinted with elemental abundances reflecting the unusually large dust concentrations of their formation sites. 

Conditions for this process are most favorable in the gas disks of high-redshift galaxies, because they are primarily neutral (so collisional drag as opposed to Coulomb and Lorentz forces dominates dust dynamics, and grain ices are not suppressed by the radiation field), and because they are metal-poor, so that under ``mean'' conditions cooling and star formation (at least low-mass star formation) are inefficient. The latter means that the surviving low-mass ``relics'' of this era will be biased towards the products of the dust-concentration process we describe, even if the conditions required are rare.

\subsection{``Initial Conditions''}

Consider a (mini)-halo at pre-reionization redshifts ($z\gtrsim6$), which recently formed a first generation of stars. Such halos are predicted to contain a gaseous disk of mostly neutral gas (free electron fractions $f_{\rm ion}\sim 10^{-6}-10^{-2}$), be rapidly polluted by trace metallicity (dispersed by SNe), and be turbulent with transsonic Mach numbers ($\mathcal{M}\equiv v_{\rm turb}/c_{s}\sim1$, owing to e.g.\ galactic rotation, accretion, and gravitational instability); see e.g.\ \citealt{wise2007:protogalaxy.collapse,wise:2008.first.star.fb,greif:turb.pop3.stars,muratov:2013.popIII.star.feedback,pawlik:2013.rad.feedback.first.stars}.

Let us make the {\em ansatz} that a non-negligible fraction of the metal mass ($\langle Z \rangle \equiv M_{\rm metal}/M_{\rm gas} = \langle \rho_{\rm metals} \rangle / \langle \rho_{\rm gas} \rangle$) is in dust ($Z_{d} \equiv \rho_{\rm dust}/\rho_{\rm gas}$). In both observations and theory \citep[e.g.][]{mathis:1977.grain.sizes,lidraine:2001.dust.model.update,draine:2003.dust.review,gordon:2003.large.variations.extinction.curves.in.lmc.smc.mw.sightlines,draine:2007.pah.model.update,de-marchi:2014.30.dor.large.micron.sized.grains.needed}, most of the grain mass is always in the largest grains, for which we define the size $\Rgrain\equiv \Rgrain_\mu\,\mu\,{\rm m}$ (typical $\Rgrain_{\mu} \sim 0.1-10$; \citealt{grun:1993.interstellar.grains.large.sizes.collected,landgraf:2000.very.large.grains.from.interstellar.seen.by.spacecraft,witt:2001.xr.halos.imply.large.dust.grains,altobelli:2007.cassini.confirms.large.microns.sized.interstellar.dust.grains,goldsmith:2008.taurus.gmc.mapping,poppe:2010.new.horizons.confirms.ulysses.large.dust.measurements,schnee:2014.mm.sized.grains.in.star.forming.regions}).

\subsection{Dust-to-Gas Fluctuations: Physics}

Large ($>$\AA-sized) dust grains do not form a fluid, but behave (in neutral gas) as aerodynamic particles\footnote{Following \citet{elmegreen:1979.charged.grain.diffusion.gmcs,draine.salpeter:ism.dust.dynamics,draine:1987.grain.charging}, we expect dust-dust collisions, Coulomb interactions, and Lorentz forces to be sub-dominant in the dust momentum equation by factors $\sim Z_{d}\,(1+\mathcal{M}^{2})^{-1/2} \ll 1$, $\sim 6\,f_{\rm ion} \ll 1$, and $\sim 0.1\,\tilde{B}\,T_{100\,K}\,\Rgrain_{\mu}^{-1}\,n_{10}^{-1/2}\ll 1$ (where $\tilde{B}$ is the ratio of the magnetic field strength to equipartition, and $n_{10}=n_{\rm gas}/10\,{\rm cm^{-3}}$), respectively. Even assuming no extinction, radiation pressure only dominates well inside of the Stromgren spheres of massive stars ($R_{\rm RP}/R_{\rm Stromgren}\approx 0.1\,(n_{10}\,L_{\ast}/10^{4}\,L_{\sun})^{1/6}$, where $R_{\rm RP}$ is the distance from a star within which radiation pressure dominates over drag). For essentially all reasonable models (discussed in the text), grain formation/destruction timescales in stars are much longer than local gas dynamical times, so do not alter the relevant instabilities. And we assume the Stokes limit for drag, trivially valid for $\Rgrain \lesssim 10^{13}\,{\rm cm}$.} which feel a drag: $D{\bf v}_{\rm dust}/Dt = -({\bf v}_{\rm dust}-{\bf v}_{\rm gas}) / \tstop$ where the drag or ``stopping'' time
\begin{align}
\tstop &\equiv \frac{\rhointernal\,\Rgrain}{\rhogas\,\cs} \approx 1.6\times10^{7}\,{\rm yr}\,\left(\frac{n_{\rm gas}}{{\rm cm}^{-3}}\right)^{-1}\left( \frac{T_{\rm gas}}{1000\,K}\right)^{-1/2}\Rgrain_\mu
\end{align}
where $\rhointernal\approx 2.4\,{\rm g\,cm^{-3}}$ is the {\em internal} grain density \citep{weingartner:2001.dust.size.distrib}, and $\rhogas$, $c_{s}$ the gas density/sound speed.

Because of this partial coupling, grains can experience large, coherent density fluctuations {\em relative to the gas}. This occurs on spatial scales $\sim R$ on which there are velocity structures (e.g.\ turbulent eddies) with characteristic timescales ``resonant'' with $\tstop$ \citep[i.e.\ $\tstop\sim|\Teddy(R)|\equiv R/\langle v_{t}^{2}(R) \rangle^{1/2}$;][]{cuzzi:2001.grain.concentration.chondrules,hogan:2007.grain.clustering.cascade.model,bec:2009.caustics.intermittency.key.to.largegrain.clustering,olla:2010.grain.preferential.concentration.randomfield.notes,hopkins:2013.grain.clustering}. If $|\Teddy|\ll \tstop$, grains simply ``pass through'' structures without significant perturbation; if $|\Teddy|\gg \tstop$, grains are well-entrained (move with the gas). In a rotating disk, the most dramatic fluctuations appear when $\tstop\sim \Omega^{-1}$ ($\Omega$ is the orbital frequency, $\Omega(r_{\rm disk}) = V_{\rm c}/r_{\rm disk} \sim (G\,M_{\rm disk}/r^{3}_{\rm disk})^{1/2}$; \citealt{bracco:1999.keplerian.largescale.grain.density.sims,johansen:2007.streaming.instab.sims,carballido:2008.grain.streaming.instab.sims,bai:2010.grain.streaming.sims.test,dittrich:2013.grain.clustering.mri.disk.sims,jalali:2013.streaming.instability.largescales,hopkins:2014.pebble.pile.formation}). This scale corresponds to ``resonance'' with both the driving scale of turbulence (where most of the power is) and orbital shear (leading to new phenomena like global ``vortex traps,'' ``pressure traps,'' and zonal flows; see references above). These are well-studied in the proto-planetary disk literature, and are believed to play a critical role in planet formation. We therefore define the dimensionless ``Stokes number'' 
\begin{align}
\label{eqn:taustop} \taustop &\equiv \tstop\,\Omega = \frac{2\,\rhointernal\,\Rgrain}{\Sigma_{\rm gas}}\,\frac{\sigma}{\cs} \approx 0.23\,\Rgrain_\mu\,(1+\mathcal{M}^{2})^{\frac{1}{2}}\left( \frac{10\,M_{\sun}\,{\rm pc^{-2}}}{\Sigma_{\rm gas}} \right)
\end{align}
Here $\Sigma_{\rm gas} \approx 2\,\langle \rhogas \rangle\,H$ is the gas disk surface density ($H = \sigma\,\Omega^{-1}$ its scale-height, with $\sigma^{2}=c_{s}^{2} + v_{\rm turb}^{2}(H)$). 

A value of $\taustop\sim1$ predicts maximal fluctuations in the dust-to-gas ratio, on large size scales $\sim H$. Several groups have studied a case very similar to that of interest here: $\taustop\sim0.01-1$ in a primarily neutral disk with sufficiently low $Z_{d}$ such that ``back reaction'' (which occurs when $Z_{d}\gg 1$) can be neglected. These studies \citep[e.g.][]{johansen:2007.streaming.instab.sims,
hogan:2007.grain.clustering.cascade.model,bai:2010.streaming.instability,pan:2011.grain.clustering.midstokes.sims,dittrich:2013.grain.clustering.mri.disk.sims,hendrix:2014.dust.kh.instability} find an approximately log-normal distribution of dust-to-gas ratios, with rms scatter on large ($\sim H$) scales of $\sim (0.6,\,0.9,\,1.2)\,$dex for $\taustop=(0.1,\,0.3,\,1.0)$.\footnote{In contrast, for small nm-scale dust with $\taustop \sim 10^{-3}$, local fluctuations in the dust-to-gas ratio are predicted to occur, but on scales $R\sim 10^{-5}\,H$, far smaller than the scales of star formation, and so are averaged-out on larger scales. Also for these grains, Lorentz forces are likely dominant over collisional drag.} However, these studies were focused on proto-planetary disks and considered incompressible, slowly-cooling (adiabatic) gas with small Mach numbers.

\begin{figure*}
 \begin{tabular}{ccc}
  \includegraphics[width=0.67\columnwidth]{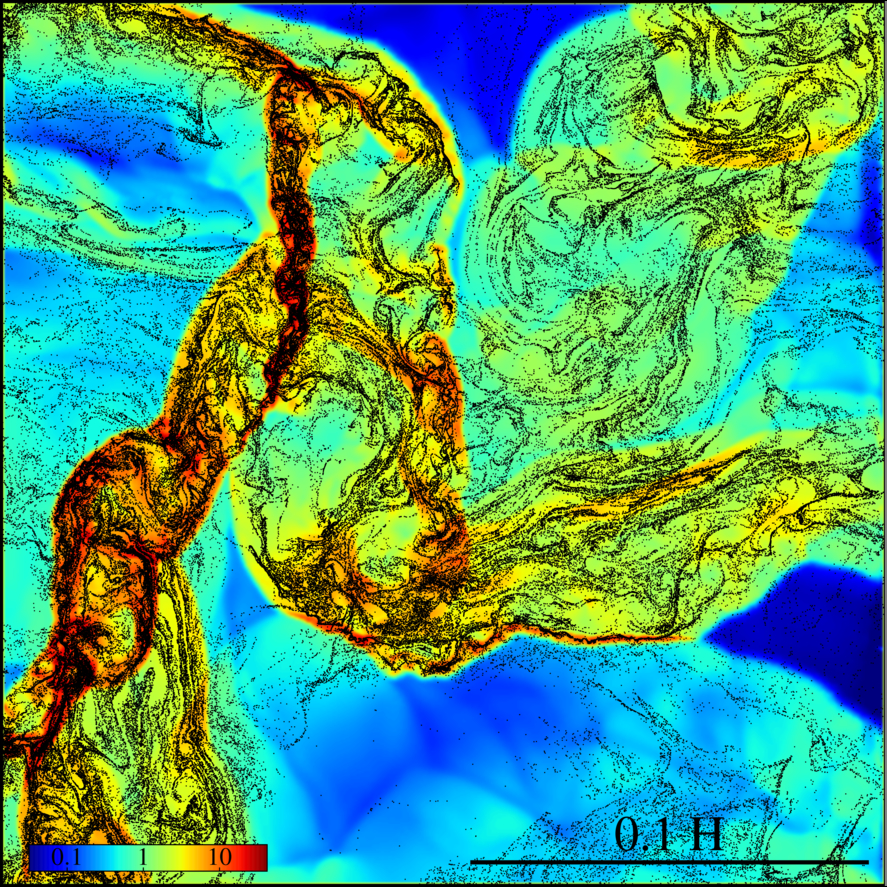} 
  \includegraphics[width=0.67\columnwidth]{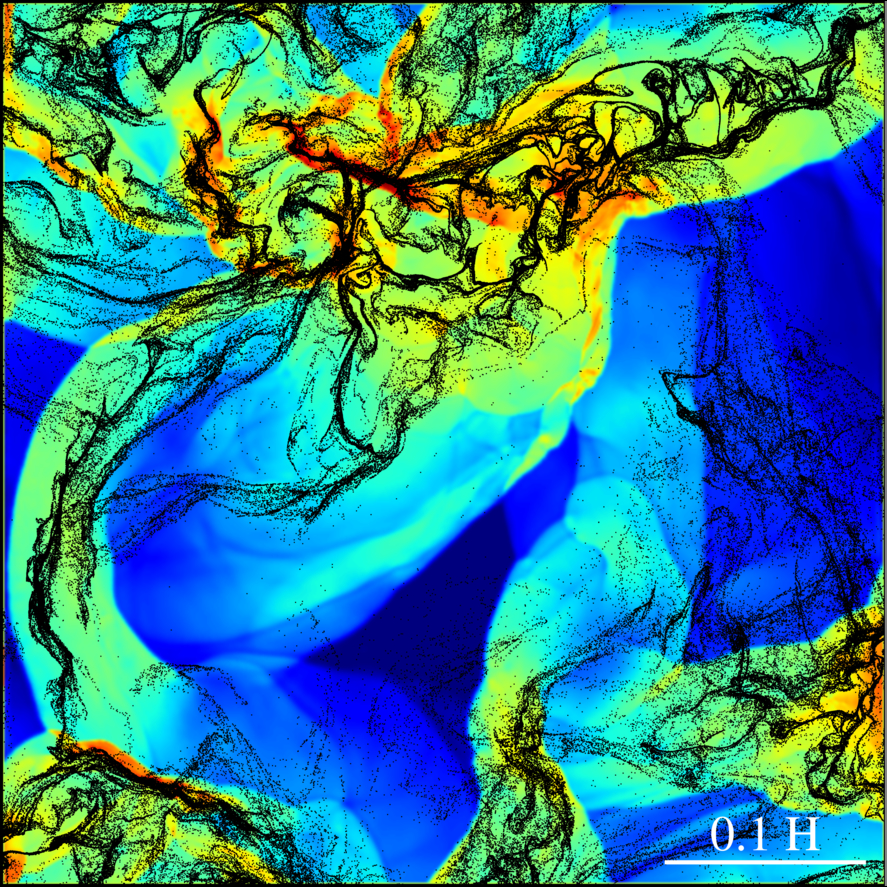} 
  \includegraphics[width=0.67\columnwidth]{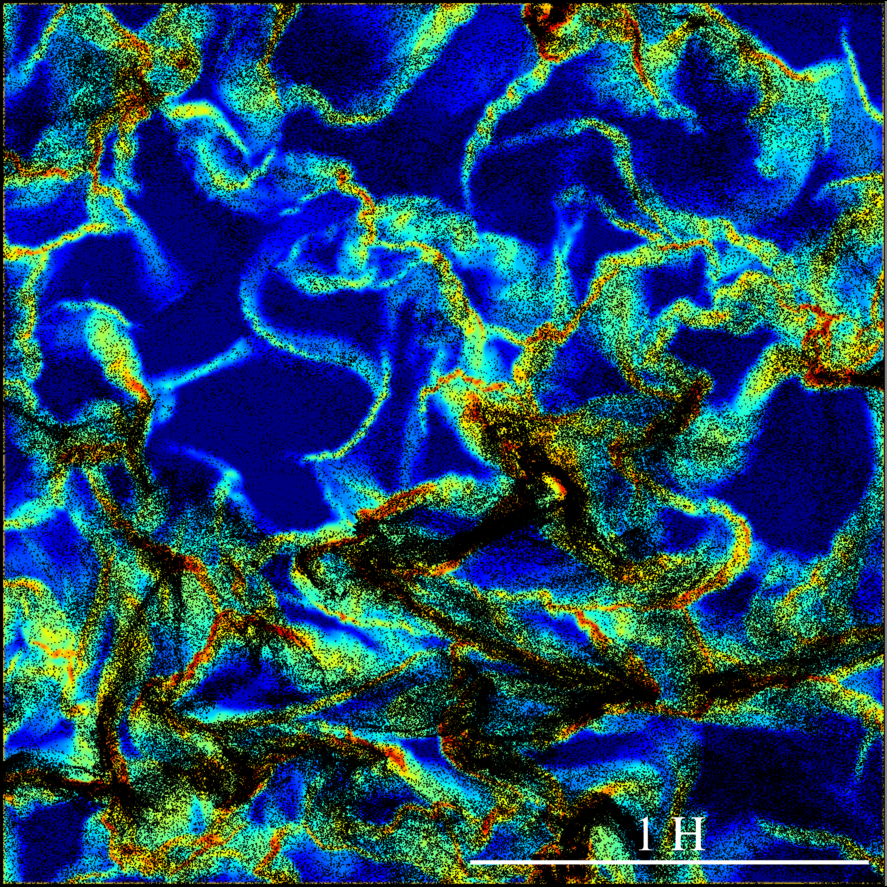} 
 \end{tabular}
    \caption{Simulations of aerodynamic dust grains in a shearing, magnetized, turbulent galactic disk, with $\langle\taustop\rangle=0.01$ ({\em left}), $\langle\taustop\rangle=0.1$ ({\em middle}), $\langle\taustop\rangle=1$ ({\em right}). Colors show gas density relative to mean ($\rho/\langle \rho \rangle$; see colorbar); black points show dust particles. We show a sub-volume of each simulation; scale is shown relative to the disk scale height $H$.     
    The simulations follow grains of a given size including gravitational forces and gas drag, in a trans-sonically MHD-turbulent (Mach numbers $\mathcal{M}\sim 2$) isothermal gas disk, representative of the neutral ISM in a high-redshift galaxy. The range of $\taustop$ corresponds to grain sizes $\sim 0.01-10\,\mu$m, depending on the disk properties (Eq.~\ref{eqn:taustop}). Dust grains clearly exhibit strong clustering, with large fluctuations in the dust-to-gas ratio. Smaller grains cluster on smaller scales; for $\taustop=0.01$ this corresponds to sub-GMC (core) scales, for $\taustop=1$ the corresponds to super-GMC scales. If the disk is metal-poor, the dust-enhanced, high-density regions may be preferentially able to form low-mass stars.
        \vspacerpostplot 
    \label{fig:turb.images}}
\end{figure*}

\begin{figure}
    \plotonesize{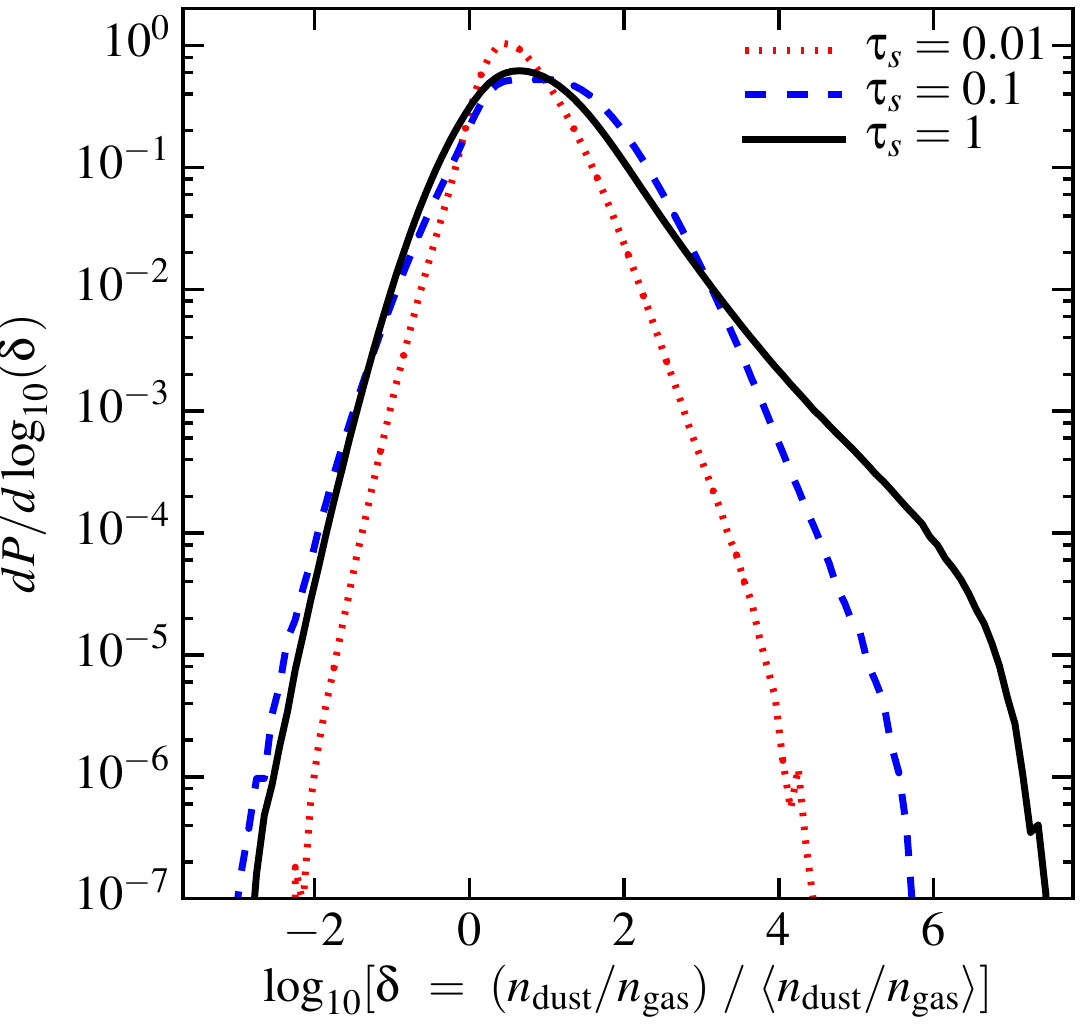}{0.9}
    \caption{Predicted dust-to-gas ratio (relative to the mean) in the simulations in Fig.~\ref{fig:turb.images}. The results are time-averaged in each case after the first few dynamical times, when the simulations reach a statistical steady-state. In all cases, there are large fluctuations in the dust-to-gas ratio as evident in Fig.~\ref{fig:turb.images}. The enhancements in the local dust density can, for large grains with $\taustop \gtrsim 0.01$, reach factors $\sim 10^{4}$.
        \vspacerpostplot 
    \label{fig:dusttogas.pdf}}
\end{figure}

\subsection{Dust-to-Gas Fluctuations: Simulations}

In \citet{hopkins.lee}, we consider a suite of simulations of aerodynamic grains, in compressible, rapidly-cooling (isothermal), magnetized, super-sonic turbulence appropriate for galactic disks. For this study, we extend that work by considering the identical setup with a different set of initial conditions (e.g.\ Mach numbers, grain sizes, etc.) chosen to reflect the parameters of interest here. Details of the simulations and extensive discussion of the relevant instabilities are in \citet{hopkins.lee}; for a summary of the specific runs shown here, see Appendix~\ref{sec:sims}. 

The simulations here are 2D (thin-disk) simulations using the shearing-sheet approximation (i.e.\ ``zooming in'' on a patch of a rotating gas disk in a constant-circular velocity potential), and solve the coupled equations of gravity, magnetohydrodynamics, and grain drag forces (with the full equations valid for compressible gases in both sub and super-sonic limits). They include the effects of the disk gravity, dust drag, hydrodynamics, magnetic fields, and turbulence, which is ``stirred'' in the large-eddy approximation to produce quasi-steady-state Mach numbers $\mathcal{M}\sim 1-2$. The grain properties are specified by $\rhointernal\,\Rgrain$ or equivalently the average value $\langle \taustop \rangle \equiv \rhointernal\,\Rgrain\,\Omega/(\langle \rhogas \rangle \langle c_{s} \rangle)$ (since $\rho$ and $c_{s}$ can vary locally in a simulation).  

Fig.~\ref{fig:turb.images} shows images of the fully-developed turbulent dust+gas disks for $\langle \taustop \rangle = 0.01,\,0.1,\,1$. It is obvious upon visual inspection that there are large dust-to-gas ratio fluctuations, and that the characteristic size-scale of fluctuations increases with $\taustop$. With $\langle \taustop\rangle\sim0.01$, there are large fluctuations from clump-to-clump (i.e.\ star forming core-to-core) within GMC-like complexes; with $\langle \taustop \rangle \sim 1$, the fluctuations are complex-to-complex. Fig.~\ref{fig:dusttogas.pdf} quantifies this, by plotting the time-averaged distribution of dust-to-gas ratios measured in each simulation. We find, consistent with \citet{hopkins.lee}, a broad distribution of local dust-to-gas ratios, even in the very high-density gas, with a logarithmic scatter of $\sim0.3-0.6\,$dex, depending on $\taustop$ and the local gas density. This implies fluctuations as large as $\gg 100$ may, in fact, occur in real galaxies, on spatial scales comparable to or larger than the scales which will collapse to form stars (i.e.\ dense cores). We show further predictions from these simulations below.

\subsection{Consequences for Star Formation}

In turbulent proto-galaxies, self-gravitating atomic/molecular clouds are constantly forming on scales $\sim 0.01-1\,H$; even before enrichment by the first (Pop III) stars, these clouds form a broad spectrum of self-gravitating sub-structures, some of which which eventually collapse to form stars \citep{barkana:reionization.review,wise2007:protogalaxy.collapse,wise:2008.first.star.fb,greif:turb.pop3.stars,pawlik:2013.rad.feedback.first.stars}. In our simulations, these randomly sample the dust-to-gas ratio fluctuations essentially independent of gas density (this is shown explicitly in \citealt{hopkins.2016:dust.gas.molecular.cloud.dynamics.sims}, where a series of scale-free simulations of pure aerodynamic dust dynamics are used to compare the variance in the dust-to-gas ratio distribution as a function of local density over $\sim 10$ decades in density).  \citet{hopkins:frag.theory} noted that any self-gravitating structure, by definition, must have a collapse time shorter than the eddy turnover time on the same scale, which is the coherence time of the dust-to-gas ratio fluctuations on that scale. Therefore, whatever fluctuations in dust-to-gas ratio appears on cloud scales $\sim R_{\rm cloud}$ when a cloud crosses the critical density to collapse are ``captured'' and conserved. Presumably, the captured grains will eventually shatter in collisions as their relative velocity increases while the core contracts; this will re-populate small grains and return mass to gas-phase metals, until they are eventually incorporated into the star \citep[][]{hirashita:2010.grain.shattering.to.make.small.grains}.

\subsubsection{Promoted Star Formation}

Various authors have argued that there is a critical minimum total metal and/or dust abundance $Z_{\rm crit}\sim 10^{-5}-10^{-3}\,Z_{\sun}$, above which cooling is efficient, hence cold clouds can easily collapse and fragment down to stellar and sub-stellar masses (which, unlike Pop III stars,\footnote{Assuming, of course, the ``conventional wisdom'' (but still unproven assumption) that Pop III stars are exclusively high-mass with short lifetimes.} could survive to the present-day), and star formation is ``normal'' \citep[see e.g.][]{chiaki:2014.critical.dust.abundance.for.cooling,ji:2014.si.dust.cooling.threshold.for.early.stars,ji:2014.silicate.dust.cooling.for.metal.poor.star.criterion.and.tests}.\footnote{Although others have argued that it may be possible to form at least some low-mass stars at any non-zero dust abundance, they agree that the probability of low-mass star formation increases rapidly with the dust abundance above $Z_{\rm crit}\sim 10^{-5}\,Z_{\sun}$ \citep{dopke.2013:fragmentation.all.dust.levels.but.enhanced.with.crit.dust}.} 

Therefore, even if large positive enhancements in $Z_{d}$ are rare, these may be special under high-redshift conditions. Even if $\langle Z \rangle \ll Z_{\rm crit}$, it is possible to form regions where, locally, $Z_{d}\gtrsim Z_{\rm crit}$. Because factor $\gg100$ fluctuations are possible in $Z_{d}$, if $\gtrsim10\%$ of the metals are in large grains, this implies that galaxies with {\em average} metallicities as low as $\langle Z \rangle_{\rm min} \sim 0.01\,Z_{\rm crit} \sim 10^{-7}-10^{-5}\,Z_{\sun}$ might be able to produce regions with local $Z_{d}>Z_{\rm crit}$ and form at least some low-mass stars. This extremely low $\langle Z \rangle_{\rm min}$ corresponds to halos with masses $M_{\rm halo}\lesssim 10^{8.5}-10^{10}\,M_{\sun}$ (atomic-cooling and smaller halos) enriched by at least one core-collapse event. 

This actually makes it more likely that we might observe signatures of special SNe in metal abundance patterns in the long-lived low-mass stars of the next generation. Otherwise, without ``promoted'' star formation, enriching the gas in a $\sim 10^{10}\,M_{\sun}$ halo to a uniform $Z_{\rm crit} \sim 10^{-3}\,Z_{\sun}$ would require $\sim 10^{4}$ SNe, implying that the abundances of (the most) metal-poor stars should reflect IMF-averaged yields (consistent with our argument from the mass-metallicity relation in \S~\ref{sec:intro}). 

However, we might ask whether large grains {\em alone} are a sufficient coolant. The detailed calculations above have shown that dust cooling {alone} is indeed sufficient -- in fact it is significantly more efficient than gas-phase metal cooling -- at producing fragmentation down to the sub-stellar mass scale (the minimum estimated dust-phase metallicity for fragmentation $\sim 10^{-5}\,Z_{\sun}$, while for gas-phase it is $\sim 10^{-3}\,Z_{\sun}$). \citet{schneider:2006.cloud.frag.with.dust,schneider:2012.critical.dust.abundance.for.low.mass.sf,schneider:2012.crit.dust.metallicity.specific.stars,klessen12:metal.poor.star.dust.frag.limits,nozawa.2012:dust.growth.during.collapse.promoting.fragmentation,dopke.2013:fragmentation.all.dust.levels.but.enhanced.with.crit.dust,meece.2014:dust.threshold.fragmentation.sf} and others have shown that efficient fragmentation to sub-solar masses requires the cooling time from dust $t_{\rm cool}$ be comparable (or shorter than) to the dynamical (collapse) time $t_{\rm dyn} \sim 1/\sqrt{G\,\rho}$ of a cloud by the time it reaches densities $n\gtrsim 10^{10}-10^{12}\,{\rm cm^{-3}}$ (at much higher densities, the cloud becomes too optically thick to cool, regardless of the dust content; at much lower densities, turbulence can easily generate density fluctuations and create new gravitationally unstable regions, so the details of cooling are less important). This will drop the local Jeans mass to $\lesssim 0.1\,M_{\sun}$. The gas cooling time (via large dust grains) is approximately $t_{\rm cool} \sim 1/(n_{\rm dust}\,\pi\,a^{2}\,\delta v_{\rm dust-gas}) \sim (\rhointernal\,a)/(Z_{d}\,\rho_{\rm gas}\,\delta v) \sim  \tstop/Z_{d}$, so for sufficiently large $Z_{d}$, this is shorter than $t_{\rm dyn}$. However, most of the calculations of the critical dust abundance assume a normal grain size spectrum; if we depend only on large grains (for which the cooling is less efficient), the critical dust-phase-metallicity would naively be larger. 

However, \citet{nozawa.2012:dust.growth.during.collapse.promoting.fragmentation} allow for arbitrary initial grain size distributions, and show that successful fragmentation does not depend on the initial grain size within the physical range we consider. Very crudely, we can illustrate this with the following argument: we can convert large grains to small (re-populating the size distribution) via collisions and shattering, and so increase the cooling rate (once bound in a contracting core, the large grains are expected to shatter or at least erode easily in collisions; \citealt{hirashita:2010.grain.shattering.to.make.small.grains}). However this timescale is $\sim 1/(n_{\rm dust}\,\pi\,a^{2}\,\delta v_{\rm dust-dust})$, very similar to the direct cooling time -- so it does not matter whether we cool directly off large grains or shatter/erode grains first and cool off small grains/gas. Detailed calculations by \citet{chiaki:2014.critical.dust.abundance.for.cooling,chiaki:2015.particle.splitting.truelove.criterion.not.correct.for.lagrangian.codes} allowing for grain growth during collapse find consistent results.

In either case, then, if we assume $\delta v$ tracks virial motions (or the sound speed, which should be similar in the limit where cooling is inefficient), then we obtain $t_{\rm cool}\lesssim t_{\rm dyn}$ when $Z_{d} \gtrsim 10^{-4}\,Z_{\sun}\,a_{\mu}\,(M_{\rm core}/100\,M_{\sun})^{-1/3}\,(n_{\rm gas}/10^{10}\,{\rm cm^{-3}})^{-2/3}$. So, near the mean densities of the galaxy, the large grains have no appreciable effect on cooling (but recall, turbulence dominates in this regime). But as the cloud contracts to higher densities, $t_{\rm cool}/t_{\rm dyn}$ decreases ($\propto n_{\rm gas}^{-2/3}$), until fragmentation occurs around the relevant densities ($n_{\rm gas}\lesssim 10^{10}\,{\rm cm^{-3}}$ for $Z_{d}\gtrsim 10^{-4}\,Z_{\sun}\,a_{\mu}$). For detailed calculations, see the references above.

Indeed, there is growing observational evidence favoring the dust-cooling limit as the relevant limit on the stellar abundances of extremely metal-poor stars \citep{klessen12:metal.poor.star.dust.frag.limits,debennassuti.2014:evidence.for.dust.cooling.transition.to.lowmass.sf,ji:2014.silicate.dust.cooling.for.metal.poor.star.criterion.and.tests}. This is another suggestion both that there is dust in extremely metal-poor environments, and that dust is the critical physical enabler of low-mass star formation.

If star formation occurs as a consequence of dust-enhancement, it may in turn lead to a bias in the stellar abundances observed in the relic stars, since only regions where $Z_{d} \gg \langle Z_{d} \rangle$ will form low-mass stars. We consider this below.

\subsubsection{Dust Grain Abundance Variations}

Dust-to-gas variations in cores translate to metal abundances: if a fraction $f\grain$ of a species $i$ is condensed in large grains, then a local fluctuation $\sim \delta = Z_{d}/\langle Z_{d} \rangle$ translates to a metal abundance $Z_{i} = (1-f\grain + \delta\,f\grain)\,\langle Z_{i} \rangle$. Because the grain density fluctuations are stochastic, a natural prediction of the scenario here is {\em variation} in the metal abundances produced by concentrations of large grains.

What might we expect? Although the properties of grains (e.g.\ sizes and composition) varies significantly between regions of the Milky Way and nearby galaxies \citep[see][for a review]{draine:2003.dust.review}, there are some robust conclusions suggested by many observations. The large grains containing the most mass have sizes $\sim 0.1-100\,\mu$m; their cores appear to be a mix of nearly pure-graphite ({\small C}) carbonaceous grains, and either olivine ({\small Mg$_{2}$SiO$_{4}$}) or pyroxene ({\small MgSiO$_{3}$}) silicates \citep[e.g.][]{kemper:2004.silicates.are.amorphous.and.olivine.pyroxene}.\footnote{Interestingly, observations indicate that the large silicates are iron-poor; so olivine {\small MgFeSiO$_{4}$} is either not present or, more likely, is a large component only in the small-grain population \citep{tielens:silicate.dust.composition,sofia:interstellar.abundances.and.dust,molster.2002:silicate.dust.composition,molster2003:silicate.dust.iron.poor}.} Observed ice mantles are a mix of {\small H$_{2}$O} (the dominant component), {\small CO$_{2}$}, {\small CH$_{3}$OH}, {\small NH$_{3}$}, {\small CH$_{4}$}, {\small H$_{2}$CO}, {\small CO}, and {\small XCN} (with {\small X} unknown), such that the average mass fraction is about $(0.65-0.73,\,0.12-0.15,\,0.04-0.12,\,0.08-0.1,\,<0.01)$ for ({\small O,\,C,\,N,\,H,\,X}) \citep{gerakines.1999:dust.ice.composition,gibb.2000:dust.ice.composition,chiar.2000:dust.ice.composition,keane.2001:dust.ice.composition,draine:2003.dust.review}.
 In general, the ice-covered grains will always be the largest (because the ice itself increases their size, and because ice mantles dramatically accelerate grain coagulation and ``sticking'' in grain collisions; \citealt{jones.1994:grain.shattering.vs.composition,jones.1996:dust.shattering.and.grain.size.distribution,hirashita.2009:dust.shattering.coagulation.turbulence}). The abundances of other species in large grains is, unfortunately, much less clear. {\small Na}, for example, is depleted onto dust throughout the ISM, but it is not clear whether this is in large grains \citep{jenkins:2009.depletion.patterns.vs.densities.in.ism.grains}. There can also be trace contributions from corundum ({\small Al$_{2}$O$_{3}$}), nitrides ({\small Si$_{3}$N$_{4}$}), and silicon carbides ({\small SiC}) at $\sim\mu$m sizes, as well as inclusions of {\small Ca} and {\small Ti} in silicates, but the relative abundance of these species (within grains) is usually low. 

For illustrative purposes, we will consider a highly-simplified reference model following the default model fitted to cold clouds in \citet{weingartner:2001.dust.size.distrib} and \citet{draine:2003.dust.review}, in which the large grains are a specified mix of graphite and olivine,\footnote{Specifically, we follow \citet{weingartner:2001.dust.size.distrib} and assume the large grain cores/mantles are either carbonaceous or silicate, where the carbonaceous grains are pure-{\small C} graphite chains, and the silicate grains are amorphous olivine ${\rm \small Mg}_{2}{\rm \small SiO}_{4}$; they neglect inclusions as a small correction. The mass fraction in silicate vs.\ carbonaceous grains as a function of size is given therein. We assume that, over the grain size range of interest, the ice mantles have uniform (mean) composition, with an unknown mass fraction relative to the cores that we will treat as a free parameter (roughly, the ice density is $\approx 0.5$ times the core density, so for spherical grains, the mass fraction can be easily translated into a spatial ``size'' of the mantle. For ice, following the collection of observations synthesized in \citet{draine:2003.dust.review}, we assume {\small H}$_{2}${\small O} is the dominant component, while ({\small CO}$_{2}$, {\small CH}$_{3}${\small OH}, {\small NH}$_{3}$, {\small CH}$_{4}$, {\small H}$_{2}${\small CO}, {\small CO}, {\small XCN}) have relative abundances $N({\rm \small X})/N({\rm \small H}_{2}\,{\rm \small O})=(0.13\pm0.02,\,0.12\pm0.06,\,0.20\pm0.05,\,0.02\pm0.01,\,0.04\pm0.02,\,0.06\pm0.03,\,0.03\pm0.01)$, which gives an approximate total mass fraction (summing over all species) of $(0.66,\,0.12,\,0.12,\,0.09,0.01)$ for (O,\,C,\,N,\,H,\,X) (``X'' here represents all other species; we take it to uniformly sample a solar-abundance-ratio distribution, but its abundance is so small as to be negligible).} with ice mantles of {\small O, C, N, H} with mass fractions $0.66,\,0.13,\,0.12,\,0.09$. With this simplification, the effects of dust on stellar metal abundances are specified by two parameters: the local dust-to-gas ratio, and the size of the ice mantles.

Of course, the mean grain composition (let alone the distribution of grain compositions) even at present day is uncertain -- our ``default'' model here is intended only to be illustrative for one specific example below. In most of what follows, we prefer to show a range of abundance ratios corresponding to the observed range in ISM dust. Even in the dominant species ({\small CNO}, {\small Mg}, {\small Si}) there can be $\sim 0.3$\,dex or larger variations in abundance ratios. Moreover, at high redshift, there have been suggestions that the dust composition may differ, e.g.\ becoming mostly silicate (as opposed to carbon) based \citep[see][and references therein]{cherchneff:2010.highz.silicate.dust}. Obviously, in the absence of ice mantles, this would imply little or no {\small C} in dust, hence this model would predict only enrichment in {\small Mg}, {\small Si}, and {\small O} -- however, in the presence of significant ice mantles, most of the predicted carbonaceous mass in the grains is actually in the ice mantle, and a slightly larger mantle could easily compensate for the absence of carbonaceous/graphite cores in depleting the gas-phase carbon onto grains.

\begin{figure}
\plotonesize{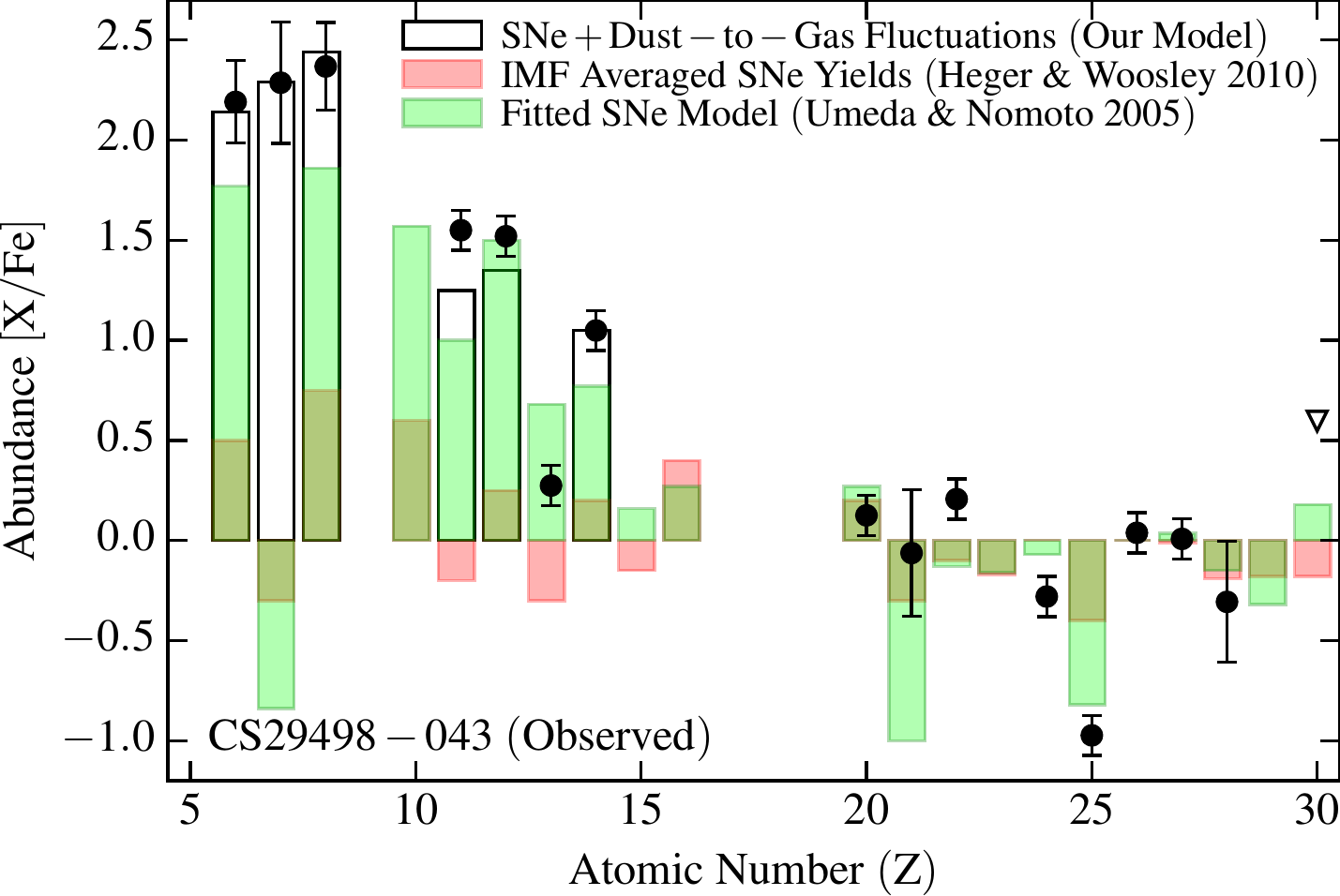}{0.99}
    \caption{Stellar abundance pattern of a typical carbon-enhanced metal-poor star (CS29498-043; {\small [Fe/H]=-3.52}; black points show data from \citealt{tominaga.2014:metal.poor.star.compilation}; arrows indicate upper limits). The IMF-averaged yields of core-collapse SNe with metal-free progenitors from \citet{heger.woosley:2010.metal.free.star.nucleosynthesis} are shown in red histograms. Adding hypernovae, freeing the progenitor metallicity, or using the IMF-averaged yields from \citet{nomoto2006:sne.yields} do not improve the agreement; clearly ``mean'' yields fail to explain this star. The blue histogram shows the best-fit abundance pattern from a single-explosion progenitor model in \citet{umeda.nomoto:2005.abundance.variation.cemp}, searching over the progenitor metallicity, mass, and explosion energy, and freely varying mixing and fallback parameters; this provides a plausible fit. However, black histograms show the prediction if we take the \citet{heger.woosley:2010.metal.free.star.nucleosynthesis} IMF-averaged yields and additionally allow for local variations in the dust-to-gas ratio according to the simulations. We take our default dust model from the text, and allow the dust-to-gas ratio and ice mantle size (two parameters) to freely vary within the range simulated, and show the best-fit result. The unusual abundances of light elements {\small C, N, O, Mg, Na, Si} could all, in principle, be explained by a local dust over-abundance in the region which was able to successfully form stars (under these metal-poor conditions), as opposed to unusual progenitor SNe.
        \vspacerpostplot 
    \label{fig:demo.abundances}}
\end{figure}

\section{Observational Signatures}

\subsection{Single-Star Abundance Patterns}

Fig.~\ref{fig:demo.abundances} considers the stellar abundances of a prototypical carbon-enhanced metal-poor star, CS29498-043 ({\small [Fe/H]}\,$=-3.52$) from \citet{tominaga.2014:metal.poor.star.compilation}.\footnote{The observations are compiled from many sources in \citet{tominaga.2014:metal.poor.star.compilation}; we plot the uncertainty-weighted average of the values there whenever multiple literature values exist.} We compare this to the IMF-averaged yields of core-collapse SNe with metal-free progenitors from \citet{heger.woosley:2010.metal.free.star.nucleosynthesis}. Although it is perfectly plausible that, at these low metallicities, a small number of SNe are responsible for the enrichment, we plot IMF-averaged values to give an idea of the mean abundance patterns expected, and/or those if the low metallicities owed to even a modest number of events mixed together and (potentially) diluted. Other models such as \citet{woosley.weaver.1995:yields} or \citet{nomoto2006:sne.yields} give similar results, even if we freely vary the progenitor metallicity (best-fit $Z\sim 0.001$) or hypernovae fraction (in the latter). Overall the IMF-averaged models do well for elements heavier than {\small Si}, as well as {\small Na} and {\small Al}.\footnote{There are some well-known modest discrepancies with the under-production of {\small Ti} and over-production of {\small Mn}; these persist even in solar-metallicity populations and probably relate to NLTE effects and other detailed modeling discrepancies.} However, there is a serious (factor $\sim20-300$) discrepancy with the light elements ({\small CNO}, {\small Mg}, {\small Si}); the models are not even qualitatively correct for these, as they predict sub-solar {\small C}, {\small N}, {\small Mg}. Adding SNe Ia components, changing progenitor metallicities, or removing the hypernovae component from \citet{nomoto2006:sne.yields} only increases these discrepancies.

Of course, it is possible to find SNe nucleosynthesis models which better fit the stellar abundances by allowing for single or arbitrary mixes of progenitor stars/explosions and then e.g.\ adjusting the progenitor abundances (although this begs the question), explosion energies, masses, density profiles, rotational support, and mixing efficiencies, and by invoking prior failed explosion/fallback episodes or ``jet ejection'' of certain species. A simple search over progenitor masses, metallicities, mixing parameters, and explosion energies in \citet{umeda.nomoto:2005.abundance.variation.cemp} gives the best-fit single-progenitor explosion model shown in Fig.~\ref{fig:demo.abundances}, based on a ``low-energy'' SNe with fallback and mixing ($25\,M_{\sun}$, $Z=0$ progenitor with explosion $E=10^{51}\,{\rm erg}$). Adding the additional degrees of freedom noted above, the best-fit model in \citet{tominaga.2014:metal.poor.star.compilation} still predicts discrepancies in {\small N}, {\small Na}, {\small Al}, and {\small Si} of $-1.8$, $-0.4$, $+0.9$, and $+0.25$\,dex, respectively, so does not represent a substantial improvement. In either case, this is of course a much better fit to the data -- and indeed represents one plausible channel for the formation of the star.

However, if we simply add our ``reference'' dust model to the IMF-averaged yields, allowing the dust-to-gas ratio ($Z_{d}/\langle Z \rangle$) and size of ice mantles to be (two) unknown parameters, we can almost-perfectly reproduce the (five) discrepant light elements!\footnote{Note that {\em if} we assume the Na is preferentially depleted onto large grains (by no means clear), we can also reproduce the observed {\small Na} abundance with dust; however, this could just as well be accounted for by metal-poor core-collapse SNe.} From the best-fit, the implied enhancement in the dust abundance in the star-forming region is $Z_{d}\sim 50-100\,\langle Z_{d} \rangle$ (well within the predicted range of our simulations). If we assume the ices surround carbonaceous grains, the best-fit abundances imply mantles of sizes $\sim 2-3$ times the cores (i.e.\ $\sim 10\%$ of {\small C} in cores, the rest in ice, implying physical mantle sizes up to $\sim 1\,\mu$m). If we assume the ices surround silicates, the mantle sizes are $\sim3$ times core sizes, and $\sim 5\%$ of the {\small Si} mass is in cores. These inferences are all consistent with common observed properties in cold, neutral regions of the local ISM \citep{witt:2001.xr.halos.imply.large.dust.grains,weingartner:2001.dust.size.distrib,costantini:2005.xray.halo.constraints.on.dust.composition,jenkins:2009.depletion.patterns.vs.densities.in.ism.grains}.\footnote{For example, the mass ratio of mantles to cores can easily be obtained if initial grain cores are produces with relatively small sizes $\sim 0.01\,\mu$m (as might be expected for SNe-produced grains after a reverse shock; see \citealt{bianchi.2007:dust.formation.sne.ejecta,nozawa.2007:dust.survival.sne.ejecta}), after which {\small CNO} was quickly depleted into ice mantles on the grains (which simple arguments suggest would produce similar-sized mantles on all grains with sizes $\sim 0.02-0.03\,\mu$m, \citealt{jones.1996:dust.shattering.and.grain.size.distribution}). Being ice-coated, the grains would then efficiently grow by sticking/coagulation \citep{jones.1994:grain.shattering.vs.composition}.}

\begin{figure}
 \begin{tabular}{c}
  \includegraphics[width=0.96\columnwidth]{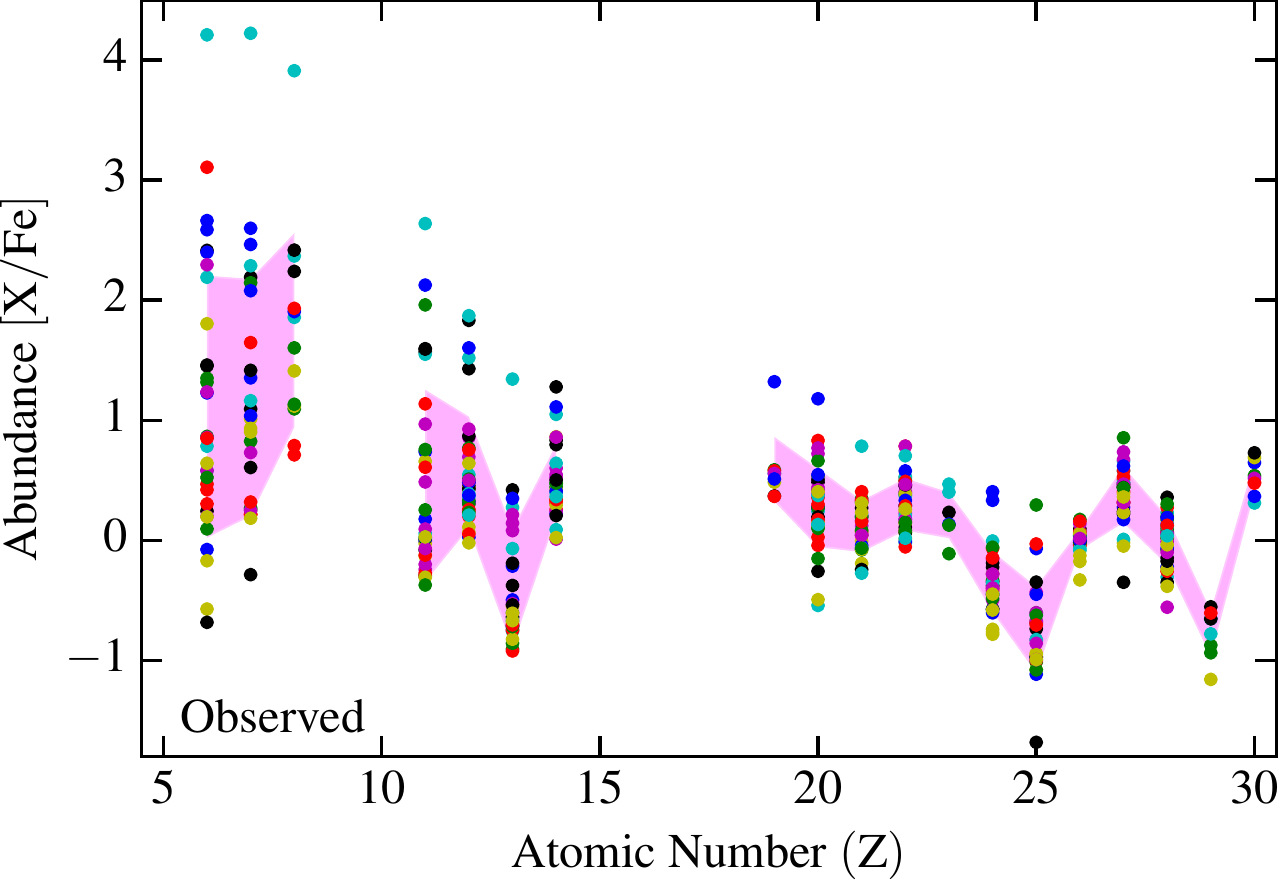} \\
  \includegraphics[width=0.97\columnwidth]{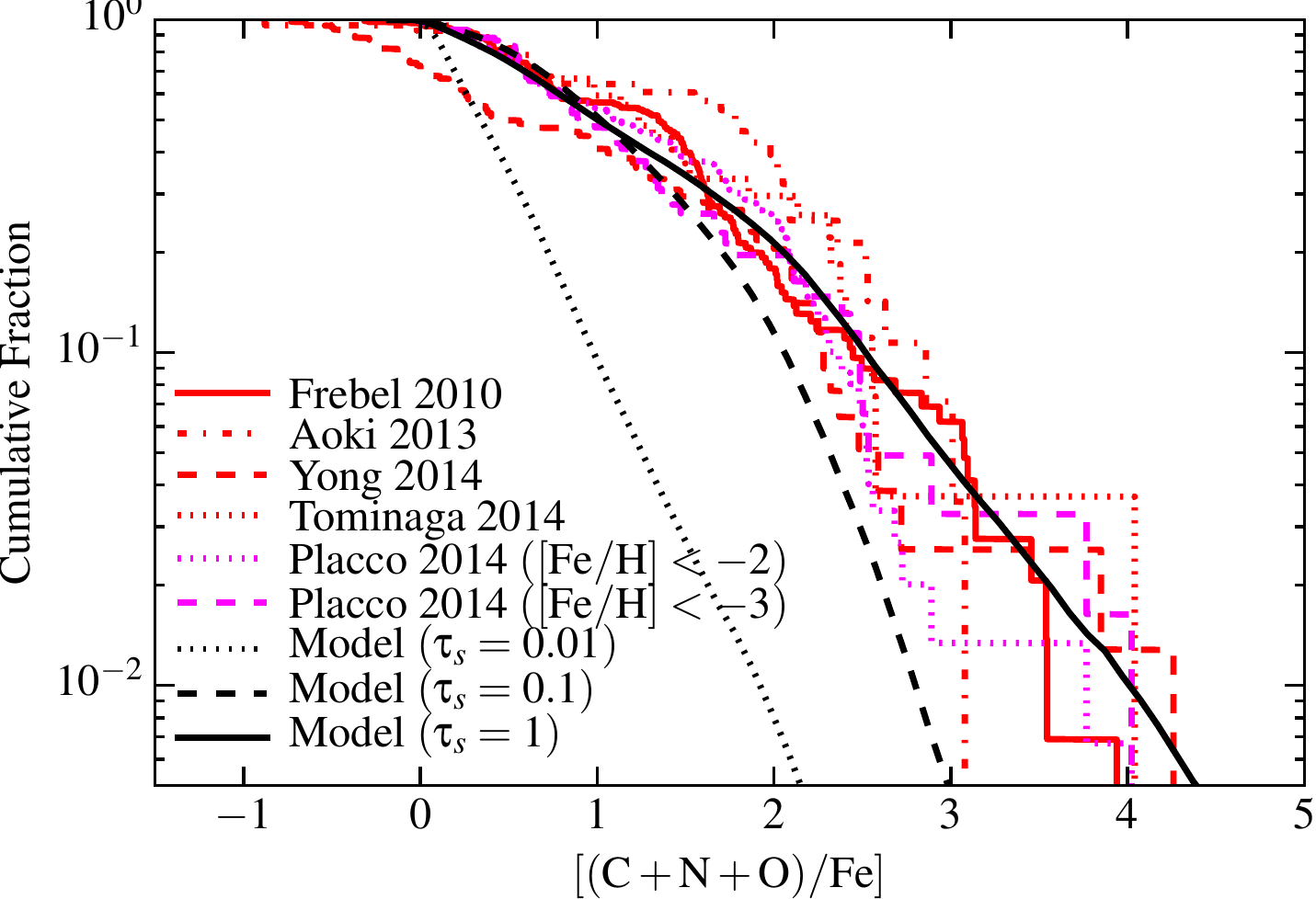}
 \end{tabular}
    \caption{{\em Top:} Stellar abundances of all stars in the \citet{tominaga.2014:metal.poor.star.compilation} sample; shaded region shows the $1\sigma$ scatter. As is well-known, the light element abundances (those which we compared in Fig.~\ref{fig:demo.abundances}) vary more dramatically than the heavy-element abundances. 
    {\em Bottom:} Distribution of {\small [C/Fe]} from the observational compilations in \citet{frebel2010:metal.poor.star.compilation,aoki2013:metal.poor.star.spectroscopy,yong2013:metal.poor.star.metallicity.distribution,tominaga.2014:metal.poor.star.compilation}. We compare the predicted distribution from simulations with our default dust model, assuming a fraction $\sim 0.5$ of the {\small C} is in large dust grains (either cores or mantles). Although there are many physical mechanisms that can lead to {\small [C/Fe]} variations, and the observational samples are incomplete, the agreement between observations and simulations is suggestive.
        \vspacerpostplot 
    \label{fig:dispersion}}
\end{figure}

\subsection{Shape of the Stellar and Dust Abundance Distributions}

Fig.~\ref{fig:dispersion} plots the stellar abundances (as Fig.~\ref{fig:demo.abundances}) for each star in the \citet{tominaga.2014:metal.poor.star.compilation} compilation. The scatter is much larger in the light elements. The scatter in {\small C} and the ice-related species, possibly due to abundance variations caused by the largest grains, is $\sim 1\,$dex. This compares to a nearly-constant $\sim 0.2\,$dex spread in all the elements from {\small Ca} to {\small Zn}, where we expect weaker dust effects. 

There are multiple possible explanations for this. However, if turbulence drives dust-density fluctuations, an approximately lognormal distribution of dust-density enhancements $\delta$ is predicted; this translates to a similar distribution in the large-grain species, but with a cutoff at low metal abundances (because some order-unity fraction of the metals are not in large grains). So in Fig.~\ref{fig:dispersion}, we collect several compilations of metal-poor stars from the literature: the \citet{tominaga.2014:metal.poor.star.compilation} sample above, as well as those in \citet{frebel2010:metal.poor.star.compilation}, \citet{aoki2013:metal.poor.star.spectroscopy}, and \citet{yong2013:metal.poor.star.metallicity.distribution}. For each, we plot the cumulative distribution of enhancements in the large-grain elements. These are highly incomplete, inhomogeneous samples, so the enhancement distribution should be regarded with great caution; but they represent nearly all the metal-poor populations known with reliable stellar abundance determinations. 

We compare this to a very simple toy model for ``promoted star formation'' based on our simulations. For each simulation, we construct $10^{6}$ random realizations; we draw a galaxy-average {\small [Fe/H]} to match the distribution of stars in the \citet{tominaga.2014:metal.poor.star.compilation} sample, apply our reference model (for IMF-averaged yields and dust), assuming $f\grain\sim1/2$ of the {\small CNO} elements are in the large grains (in ices or cores), calculate the total {\small CNO} metallicity (gas+dust) within each resolution element (above the mean density), and randomly decide whether each should form a star. Lacking any physical model for promoted star formation we simply assume the probability of star formation is proportional to the total metallicity at these low metallicities (this is arbitrary, but our intention is simply to represent the qualitative behavior we expect). Given the extreme simplicity of this model, it is remarkable how well the predictions for $\taustop\sim0.1-1$ appear to match the observed distributions.

\subsection{{\small CNO} Elements, Carbonaceous Grains, and Ices}

\begin{figure}
\plotonesize{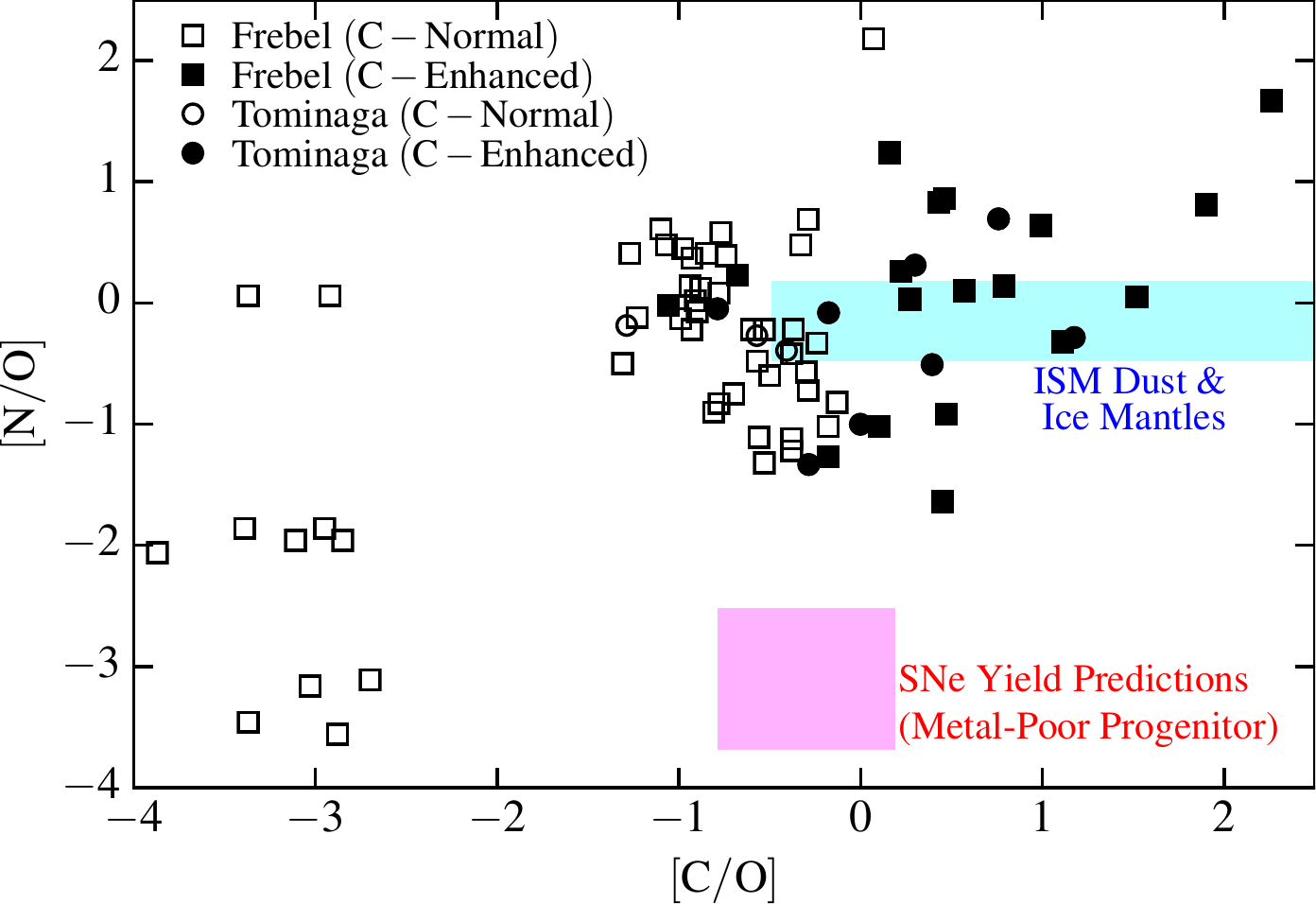}{0.95}
    \caption{Distribution of {\small CNO} elements in observed metal-poor stars, from the same compilations as Fig.~\ref{fig:dispersion}. Depending on the ratio of ice mantle to core size, large grains in the simple \citet{draine:2003.dust.review} dust model reside anywhere within the blue shaded range. The red shaded range represents the range of yields for standard SNe of any progenitor mass and explosion energy with low metallicities ($Z<0.001$) in the \citet{woosley.weaver.1995:yields} models (the \citet{nomoto2006:sne.yields} models are similar). The {\small CNO} abundances are highly degenerate with stellar evolution and pollution by winds, but the similarity of the observed abundances in the {\small C}-enhanced stars to ices, and disagreement with the low {\small N} enrichment predicted by metal-poor SNe models, is suggestive of star formation in dust-enhanced regions.
        \vspacerpostplot 
    \label{fig:cno}}
\end{figure}

%

In Fig.~\ref{fig:cno}, we consider the ratio of the {\small CNO} species observed. We caution that the ratios of the individual species involved here are subject to serious stellar evolutionary effects after the stars form, but wish to highlight that, even given this caveat, many stars have stellar abundance ratios similar to those found in ISM ices. For ice with the mean observed mixture of components in our reference model, we expect {\small [N/O]} between $-0.5$ and $+0.2$, and {\small [C/O]}$\, > -0.5$ depending on the ratio of ice mantle to underlying carbonaceous grain mass. If we consider a simple Pearson test, the enhancements {\small [C/Fe]}, {\small [N/Fe]}, and {\small [O/Fe]} are all correlated with each other ($p_{\rm uncorrelated}\ll 10^{-4}$ in both the \citealt{tominaga.2014:metal.poor.star.compilation} and \citealt{frebel2010:metal.poor.star.compilation} samples); albeit with significant scatter, as expected in the models here (as opposed to anti-correlated, if stellar evolution were dominant).

\begin{figure}
 \begin{tabular}{c}
  \includegraphics[width=0.97\columnwidth]{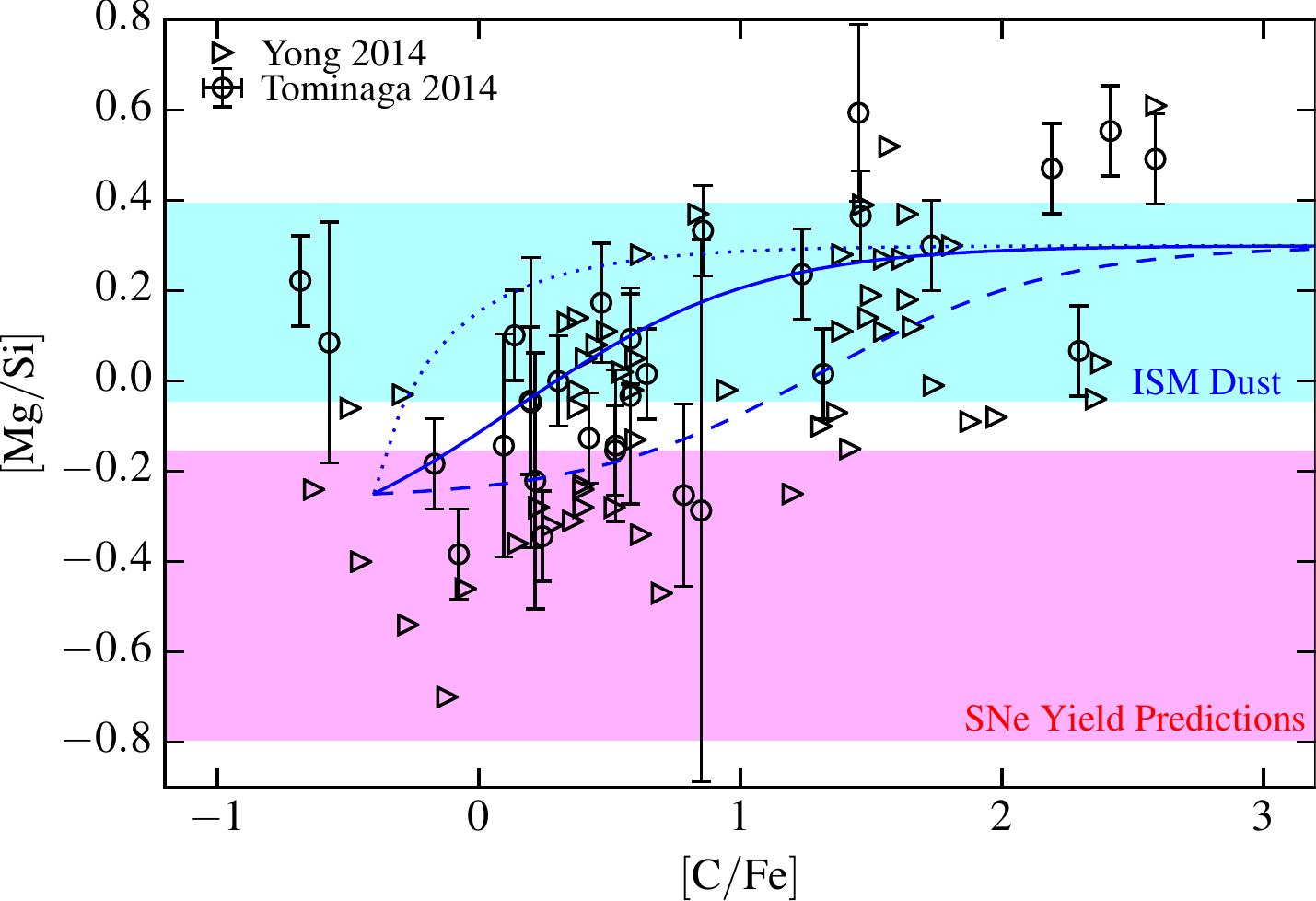} \\
  \includegraphics[width=0.97\columnwidth]{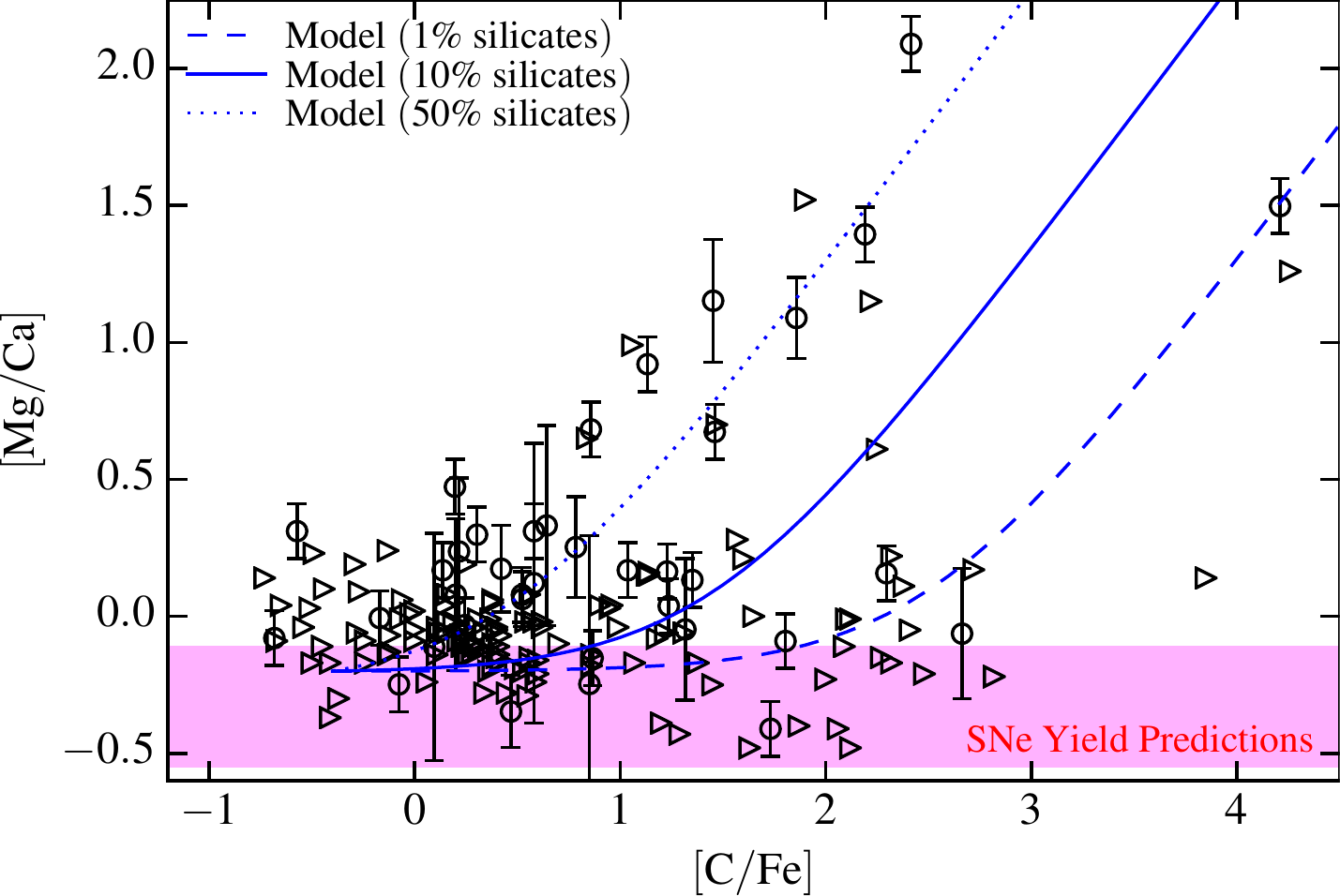} \\
  \includegraphics[width=0.97\columnwidth]{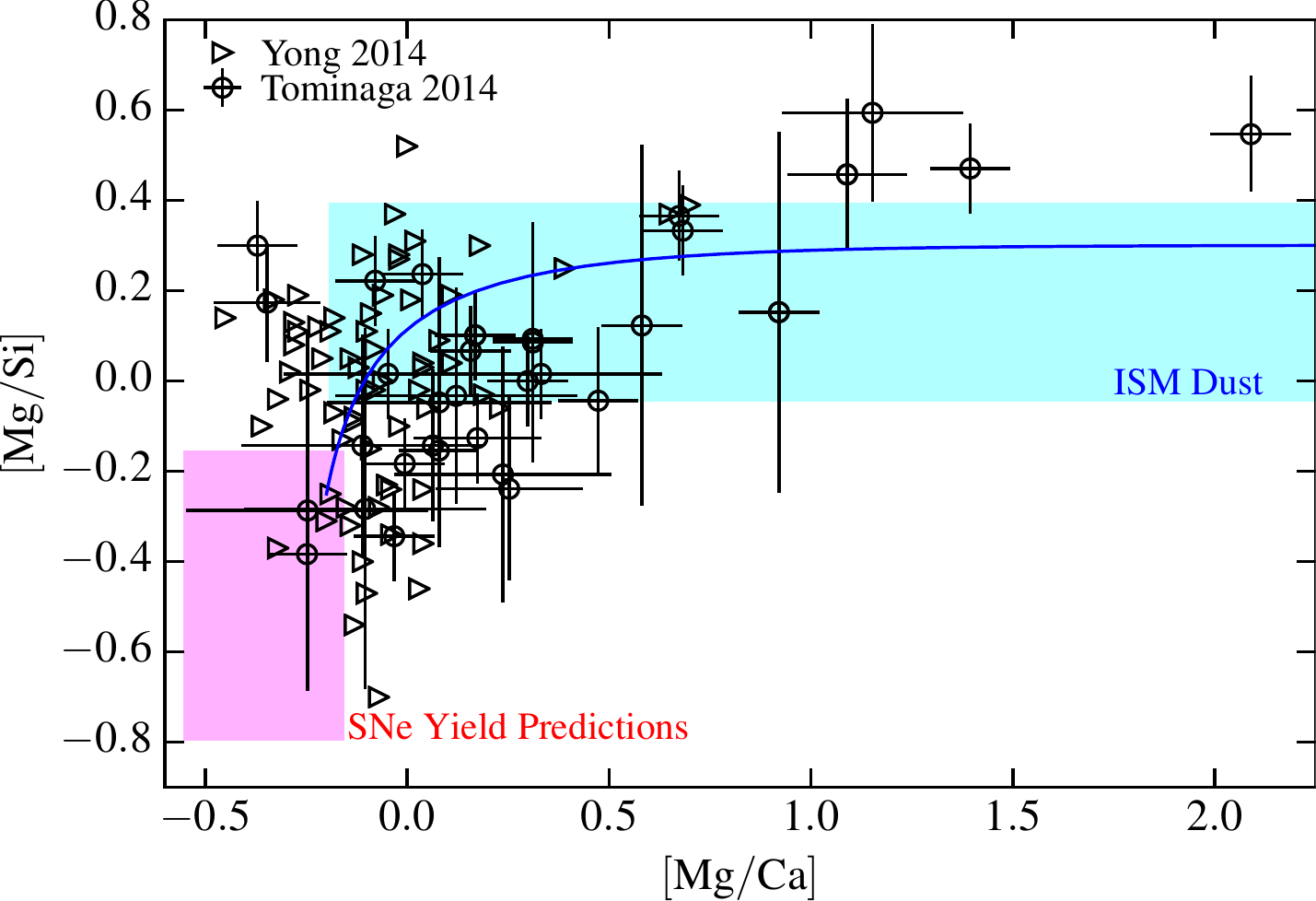}
 \end{tabular}
    \caption{{\em Top:} Observed {\small [Mg/Si]} vs.\ {\small [C/Fe]} in metal-poor stars, as Fig.~\ref{fig:cno}. {\em Middle:} Observed {\small [Mg/Ca]} vs.\ {\small [C/Fe]} in the same stars. {\em Bottom:} {\small [Mg/Si]} vs.\ {\small [Mg/Ca]}. As Fig.~\ref{fig:cno}, the range of {\small [Mg/Si]} in typical observed ISM silicate grains (at low redshift) is shown in the shaded blue range; the range of yields predicted for different individual progenitor masses and energies in metal-poor SNe is shown in shaded red. SNe models (including hypernovae and metal-rich SNe as well; see \citealt{nomoto2006:sne.yields}) tend to predict a narrow range of {\small [Mg/Ca]}. But the {\small C}-enhanced stars tend also to be enhanced in {\small [Mg/Ca]}, with similar {\small [Mg/Si]} to observed silicate grains. This is predicted if they form in dust-enhanced regions. Assuming our reference model for grains in a background with IMF-averaged yields, we predict dust-enhanced stars lie along the tracks shown as blue lines, where the track depends on the (unknown) ratio of the total mass in silicate grains to carbonaceous grains plus ices at the grain sizes which are experiencing density fluctuations. In the bottom panel, the model tracks lie on top of one another.
        \vspacerpostplot 
    \label{fig:MgSiCa}}
\end{figure}

\subsection{Distinguishing Between SNe Yields and ISM Dust with {\small Mg}, {\small Si}, and {\small Ca}}

A more robust constraint can come from the behavior of species in massive silicates. First, we should stress that, because of expected variation in the ice mantle covering and size distributions of silicate and carbonaceous grains (and the fact that grains with different $\taustop$ will, even in the same location in the same galaxy, be affected differently), we do not necessarily predict a strong correlation between {\small C} and {\small Si} enhancements. However there should be some positive trend, where stars with enhanced {\small C} are at least more likely to have enhanced variation in {\small Si}; indeed a Pearson test confirms this ($p_{\rm uncorrelated}\sim 10^{-5}$) and it is plain ``by-eye'' in both the \citet{tominaga.2014:metal.poor.star.compilation} and \citet{frebel2010:metal.poor.star.compilation} samples. 

However, where {\small Si} is enhanced (if this is because of dust) our reference model more robustly predicts a correlated enhancement in {\small Mg}, since these are coming {\em from the same grains}. Indeed, we find that, within the metal-poor population ({\small [Fe/H]}$\,<-2.5$), {\small [Si/Fe]} and {\small [Mg/Fe]} are more significantly correlated {\em with each other} ($p_{\rm uncorrelated}\sim 10^{-8}$) than with the heavier species {\small K, Ca, Sc, Ti, V, Cr, Mn, Co, Ni, Cu, Zn}, ($p_{\rm uncorrelated}\sim0.1-1$).\footnote{There are marginally significant correlations between {\small [Si/Fe]} or {\small [Mg/Fe]} and {\small [Na/Fe]} or {\small [Al/Fe]}, which could stem either from enrichment by prior SNe or dust contamination (with some amount of dust containing {\small Na} and {\small Al}, as discussed above).} Fig.~\ref{fig:MgSiCa} shows this more quantitatively. We plot {\small [Mg/Si]} and {\small [Mg/Ca]} versus {\small [C/Fe]}. If the silicate grain cores are a mix of olivine ({\small [Mg/Si]}$\,\sim +0.3$) and pyroxene ({\small [Mg/Si]}$\,\sim -0.05$), we expect {\small [Mg/Si]} in a relatively narrow range. Fig.~\ref{fig:MgSiCa} confirms this, especially for stars with {\small [C/Fe]}$\,\gtrsim 0.5$ (given measurement errors, $>90\%$ of the stars with this {\small [C/Fe]} are consistent with the predicted range in {\small [Mg/Si]}).\footnote{Note there also appear to be some, albeit few, stars which have highly-enhanced {\small Ca} at low metallicities \citep[see][]{lai:ca.rich.metal.poor.star}.}

For comparison, in burning models, the production of {\small Mg} and {\small Ca} (produced by similar processes in similar-mass stars) should be much more closely linked than {\small Mg} and {\small Si}. For pure, metal-poor, IMF-integrated ejecta, these models predict {\small [Mg/Ca]}$\,\sim -0.3$ with relatively small scatter \citep{woosley.weaver.1995:yields}. In fact, we find nearly every star is enhanced above this level, as predicted if the star-forming region was ``dust-enhanced.'' Interestingly, if we restrict to more metal-rich stars which are not light-element enhanced ({\small [Fe/H]}$\,>-2.5$, with {\small [Mg/Fe]}$\, < 0.5$), {\em then} the correlation between {\small Mg} and {\small Ca} is good. Within that population, we expect the fraction of ``dust enhanced'' stars to be small.

\section{Discussion \&\ Conclusions}

We show that, provided dust grains with sizes $\gtrsim 0.01-0.1\,\mu$m exist in high-redshift, predominantly neutral galaxies, the relatively weak coupling between the dust and gas means that these grains can experience orders-of-magnitude fluctuations in their local number density (dust-to-gas ratio). 

This means there can be efficient cooling and low-mass (Pop-II) star formation in clouds with high {\em local} dust density, even when the galaxy-average metallicity is extremely low ($\langle \log{[Z/H]} \rangle \ll -5$). In other words, second-generation star formation can begin (albeit stochastically) almost immediately after the first dust is produced (via Pop-III winds or SNe), without waiting for any galaxy-wide enrichment threshold to be reached. We refer to this mechanism as ``promoted star formation.''

This may also have significant impacts on the stellar abundance patterns of certain elements in the stars which form from these clouds. We show that this can naturally explain otherwise unusual abundance patterns in metal-poor stars, including the large carbon-enhanced (and {\small CNO}-enhanced) population, and stars with elevated, tightly correlated {\small Si} and {\small Mg} without {\small Ca} enhancement. This would explain growing observational indications of independent formation channels for the observed CEMP-no and carbon-normal metal-poor stars \citep{norris:2013.two.metal.poor.star.pops}. Compared to SNe nucleosynthesis models, better fits to the stellar abundances are found with fewer free parameters, for a substantial sub-population of stars, using a simple standard model of dust chemistry coupled to direct numerical simulations of dust dynamics. This simple ``dust-enhancement'' model also naturally predicts a quasi-lognormal distribution of abundances in good agreement with observations, and typical abundance ratios of {\small CNO} and {\small Si}, {\small Mg}, {\small Ca} which are observed but are difficult if not impossible to explain in most SNe nucleosynthesis models. 

The key theoretical assumption here is that there {\em is} dust, containing a non-negligible fraction of metals, in such low-metallicity galaxies. According to some models, this is unlikely. However, a growing body of observations (as well as newer theoretical models; see e.g.\ \citealt{hirashita:2014.dust.from.sne.in.highz.gals,mattsson:2014.dust.from.ism.growth.at.highz,marassi:dust.yield.sne.metal.poor,marassi:sne.dust.yield.sf.scenario}) indicate it may be inevitable. Absorption in gamma ray bursts, quasars, and lensed galaxies at high redshifts (many at $z\gtrsim5-7$) indicates they have ``normal'' dust-to-metal ratios, despite metallicities as low as {\small [Z/H]}$\,\sim-2.5$ \citep{cucchiara:2011.z9.grb.with.dust,zafar:2013.dust.to.metals.highz.grb.hosts.is.high,chen:2013.dust.in.lensed.gals,kuo:2013.dust.grb.afterglows,de-cia:2013.grb.host.dust.to.z4,sparre:2014.z5.grb.dust,dwek:2014.z9.6.dust.candidate.detection}. In fact observational compilations from $z\sim0-10$ argue there is little significant evolution in the dust-to-metal ratio (references above), extinction curve \citep{cucchiara:2011.z9.grb.with.dust,sparre:2014.z5.grb.dust}, maximum dust sizes (grains $>0.1\,\mu$m required at all times; \citealt{updike2011:grb.host.dust,hirashita:2012.highz.grain.size.distrib.models,kuo:2013.dust.grb.afterglows}), or the carbon/silicate ratio in the large grains (though this is more uncertain; see e.g.\ \citealt{kuo:2013.dust.grb.afterglows,dwek:2014.z9.6.dust.candidate.detection}).\footnote{It has been noted by many authors that in high-redshift galaxies and local low-metallicity galaxies like the LMC and SMC, the relative weakness of the $2175$\AA\ feature suggests a lower ratio of graphite dust to silicate dust \citep[see e.g.][]{pei92:reddening.curves,richards:red.qsos,hopkins:dust,maiolino.2004:qso.dust.sne,li2008:grb.host.dust.smc.like,perley.2010:sne.synthesized.dust.in.grbs,perley.2011:grb.dust.smc.like,updike2011:grb.host.dust,zafar:2012.2175.angstrom.bump.grbs,schady:2012.grb.no.2175.extinction.bump}. However, this only constrains carbonaceous grains with sizes $\lesssim 10\,$nm ($\Rgrain_{\mu} < 0.01$), irrelevant for our purposes; in fact in most models, if these are removed, carbon is incorporated in larger grains, enhancing the effects we propose here.}. It has also been observed that SNe can directly produce large quantities of dust (an order-unity fraction of the ejecta metals), with especially large grain sizes up to $\sim 10\,\mu$m which would survive reverse shocks -- these appear especially in SNe of types (e.g.\ IIn) which some authors have speculated occur preferentially at low metallicity \citep{gall:2014.several.micron.grains.forming.directly.in.sne,de-marchi:2014.30.dor.large.micron.sized.grains.needed,kochanek:2014.large.dust.formed.by.failed.SNe,marassi:dust.yield.sne.metal.poor}. The same may be true of early stellar winds \citep{nozawa:2014.rapid.production.of.grains.in.massive.stars.at.highz}. And finally, more and more observational data favors dust as the critical coolant enabling fragmentation and formation of the observed, low-mass extremely metal-poor stars \citep{schneider:2012.critical.dust.abundance.for.low.mass.sf,klessen12:metal.poor.star.dust.frag.limits,debennassuti.2014:evidence.for.dust.cooling.transition.to.lowmass.sf,ji:2014.silicate.dust.cooling.for.metal.poor.star.criterion.and.tests,chiaki:2014.critical.dust.abundance.for.cooling}; this suggests that dust must have been present in substantial amounts relative to gas when these stars formed.

Given the presence of dust in a neutral, modestly MHD-turbulent gas disk with ratio of dust stopping time to orbital time $\taustop\sim 1$, there are many analogies between this problem and the well-studied problem of dust dynamics in proto-stellar disks. What most importantly differentiates the galactic case is (1) the absolute scales are different, so micron-sized dust in a galactic mini-halo behaves like meter-sized boulders in a proto-planetary disk, (2) the gas is compressible, which can enhance fluctuations in the dust-to-gas ratio, and (3) self-gravity is much more important in the galactic case, while the mean metallicities are much lower, so we do not expect to form ``super-planetesimals,'' but rather to ``promote'' normal star formation via cooling in regions where the dust abundance is relatively high.

What makes high-redshift galaxies special compared to low-redshift galaxies is (1) they are primarily neutral, even at surface densities where $\taustop\sim1$. In the Milky Way, these densities would be fully ionized, so Coulomb and Lorentz forces would dominate the dust dynamics; moreover the much stronger radiation field in the Milky Way suppresses ice formation and grain coagulation. And (2) the early galaxies are metal-poor, so cooling and star formation are expected to be less efficient under ``mean'' conditions. As such, regions with unusual dust/metal abundances might be the only regions capable of low-mass star formation, and therefore the observed stellar ``relics'' of this era will be preferentially biased towards these abundances. In contrast, while we do predict that qualitatively similar processes can occur in at least some neutral regions of galaxies today (certain large molecular clouds, for example), the ``dust-enhanced'' stars would represent only a miniscule fraction of the stellar populations forming in these regions \citep[see][]{hopkins:totally.metal.stars}. 

There are many questions this raises which merit further study. The same instabilities shown here may dramatically enhance grain formation and growth in proto-galaxies; with nominal ``clumping factors'' $\langle \rho_{\rm dust}^{2} \rangle / \langle \rho_{\rm dust} \rangle^{2}$ reaching $\sim 1000$, these effects need to be incorporated into dust growth models. Our dust dynamics simulations are also being extended to global disk simulations, with explicit models for Lorentz forces (neglected here). But even these are not actual star formation simulations. However, the simple equations of dust dynamics could be incorporated into self-consistent simulations of star formation which include detailed chemistry, dust+gas cooling, self-gravity, and dynamical enrichment models. This would enable more detailed, quantitative predictions for the importance of ``promoted star formation'' in dust-enhanced regions. 

Perhaps most importantly, this motivates more detailed study and models of dust chemistry in high-redshift galaxies. For the sake of simplicity (and to make progress) we adopted a dust model calibrated to local observations, but this is almost certainly incorrect (although it provides a surprisingly good fit to observations). We believe that this is the largest uncertainty in predicting observable stellar abundance signatures from dust. Of course the elements heavier than {\small H} and {\small He} in dust must come from nucleosynthesis originally; therefore, one might expect if different gas-phase metal abundance ratios are present at high redshift, the dust will reflect this. In the limit where the large dust grains perfectly trace the gas-phase metal abundances, it obviously becomes impossible to identify the mechanism we propose here via stellar abundance patterns (although this limit is unlikely, given the range in condensation temperatures of different species). However, such mechanisms could still be critical in producing dust-to-gas variations that allow ``promoted'' star formation. 

\acknowledgments 
We thank Jessie Christiansen, Evan Kirby, and Selma de Mink for many helpful discussions during the development of this work. Support for PFH was provided by an Alfred P. Sloan Research Fellowship, NASA ATP Grant NNX14AH35G, and NSF Collaborative Research Grant \#1411920 and CAREER grant \#1455342. Numerical calculations were run on the Caltech compute cluster ``Zwicky'' (NSF MRI award \#PHY-0960291) and allocation TG-AST130039 granted by the Extreme Science and Engineering Discovery Environment (XSEDE) supported by the NSF. CC is supported by the Packard Foundation, NASA grant NNX13AI46G, and NSF grant AST-1313280.\\

\bibliography{/Users/phopkins/Dropbox/Public/ms}

\begin{appendix}

\section{Details of the Simulations}
\label{sec:sims}

\subsection{Numerical Method}

The simulations in the text are from the set in \citet{hopkins.lee}. They solve the standard equations of magnetohydrodynamics (MHD) using {\small GIZMO} \citep{hopkins:gizmo}, a mesh-free, Lagrangian finite-volume Godunov code which captures advantages of both grid-based and smoothed-particle hydrodynamics (SPH) methods. In \citet{hopkins:gizmo} and \citet{hopkins:mhd.gizmo}, we consider extensive tests of the method, and show that {\small GIZMO} agrees very well with state-of-the-art moving-mesh and grid-based adaptive-mesh refinement codes on both sub and super-sonic MHD turbulence. 

The turbulent driving routines are the same as used for these tests (see \citealt{hopkins:lagrangian.pressure.sph} and \citealt{hopkins:gizmo} for details), and follow \citet{bauer:2011.sph.vs.arepo.shocks}. The box is stirred via the same method as \citet{schmidt:2008.turb.structure.fns,federrath:2008.density.pdf.vs.forcingtype,price:2010.grid.sph.compare.turbulence}: a small range of modes are driven in Fourier space as a purely-solenoidal Ornstein-Uhlenbeck process. After a short period of initial adjustment ($\sim 1$ turbulent crossing time), this maintains a quasi-steady-state Mach number and turbulent cascade; the statistically properties of the simulations remain constant after this time. 

We adopt the well-known shearing-sheet approximation \citep[for details see e.g.][]{guan:2008.shearing.box.mri}. Here we solve the vertically-integrated equations for dust and gas in 2D (following radial/azimuthal $R$, $\phi$ coordinates), in a local frame which co-rotates with circular orbits with a frame-centered orbital frequency $\Omega$. This amounts to adopting standard shear-periodic boundary conditions, with the centrifugal and Coriolis accelerations ${\bf a} = 2\,q\,x\,\Omega^{2}\,\hat{x} + 2\,{\bf v}\times(\Omega\,\hat{z})$ (where $q\equiv-d\ln{\Omega}/d\ln{R}$). 

Following most previous studies \citep[see e.g.][]{carballido:2008.grain.streaming.instab.sims,hogan:1999.turb.concentration.sims,johansen:2007.streaming.instab.sims,johansen:2009.particle.clumping.metallicity.dependence,bai:2010.grain.streaming.sims.test,pan:2011.grain.clustering.midstokes.sims}, we represent the dust via ``super-particles'' each one of which represents an ensemble of grains, with trajectories integrated on the fly. Following \citet{draine.salpeter:ism.dust.dynamics}, the grains obey the equations of gravity along with the drag equation: 
\begin{align}
\label{eqn:grain.eom} \frac{d{\bf u}_{d}}{dt} &= -\frac{{\bf u}_{d}-{\bf u}_{\rm gas}}{t_{s}} \\ 
\label{eqn:tstop} t_{s} &\equiv \frac{\pi^{1/2}}{2\sqrt{2}}\,\left( \frac{\rhointernal\,\Rgrain}{c_{s}\,\rho_{\rm gas}} \right)\,\left( 1 + \left|\frac{3\pi^{1/2}}{8}\,\frac{{\bf u}_{d}-{\bf u}_{\rm gas}}{c_{s}} \right|^{2}\right)^{-1/2}
\end{align}
where ${\bf u}_{d}$ is the grain velocity, $d/dt$ is a Lagrangian derivative, $c_{s}$ and $\rho_{\rm gas}$ the isothermal sound speed and density of the gas, $\rhointernal$ is the internal (material) grain density, and $\Rgrain$ is the grain radius. The time-integration scheme is described in \citet{hopkins.lee}. Note that this is the proper expression for super-sonic flows, and both $c_{s}$ and $\rho_{\rm gas}$ are the values evaluated {\em at the grain position}, which can be spatially variable: this is the major difference between our study and the previous work on proto-planetary disks, which assume Mach numbers $\mathcal{M}\ll1$, and therefore drop  the velocity-dependent term in $t_{s}$ and take $\rho_{\rm gas}=$\,constant. 

\subsection{Approximations}

As noted in the text, this assumes grains are in the Stokes regime which is trivially satisfied for $\Rgrain \ll 10^{13}\,{\rm cm}$. Because the absolute metallicities $Z_{d}$ of interest are low ({\small [Z/H]}\,$\lesssim -2.5$, hence $\langle Z_{d} \rangle \lesssim 10^{-5}$), we can safely neglect back-reaction (i.e.\ the momentum loss from gas onto grains), which is only important for local $Z_{d} \gg 1$. Also as noted in the text, radiation pressure and Coulomb forces are negligible in the regime we simulate (primarily neutral gas, not in the HII region of individual stars). 

The case of Lorentz forces on dust is less clear. Adopting equipartition magnetic fields and a mean grain charge as a function of grain size estimated in \citet{draine:1987.grain.charging}, and assuming grain motion is entirely perpendicular to field lines, we noted in the text that the ratio of Lorentz to drag forces is $\sim 0.1\,T_{100\,{\rm K}}\,\Rgrain_{\mu}^{-1}\,n_{10}^{-1/2}$. So while negligible for the largest grains in high-density regions (cores) which will actually form stars, it is by no means clear that we can completely ignore Lorentz forces \citep[for a more detailed analytic comparison, see][]{yan.2004:lorentz.forces.drag.dust.ism.analytic}. Unfortunately both the grain charge and magnetic field strength are highly uncertain under the conditions of interest; we therefore do not explicitly include Lorentz forces in our simulations. In future work, however, we intend to investigate this in more detail.

To first approximation, in most studies of turbulence and star formation, the full effects of cooling can be reasonably represented by simply adopting an isothermal ($\gamma=1$) equation of state \citep[see e.g.][]{nordlund:1999.density.pdf.supersonic,ostriker:1999.density.pdf,li:2005.turb.reg.sfr,krumholz:2007.rhd.protostar.modes,veltchev:2011.frag,kritsuk:2011.mhd.turb.comparison,molina:2012.mhd.mach.dispersion.relation,hopkins:2012.intermittent.turb.density.pdfs,konstantin:mach.compressive.relation}. We do so here. Although cooling physics may be more complicated and less efficient under high-redshift conditions, the isothermal approximation is still reasonable; moreover we have considered the opposite regime of ``inefficient cooling'' and run two simulations with $\gamma=7/5$ and $\gamma=5/3$, respectively. Although the structures which form in the {\em gas} are modified (as the gas is less compressible), the {\em dust-to-gas} ratio fluctuations are qualitatively identical (presumably because these are driven by the vorticity field, not compressible gas motions). In fact the maximum dust-to-gas ratio fluctuations have slightly {\em larger} magnitude in these ``adiabatic'' cases.

Obviously, for the sake of computational expense and resolution, we simplify by restricting to 2D. However, in \citet{hopkins.lee}, we consider a suite of both 2D and 3D simulations (lower-resolution and without shear) and show that all the same conclusions apply in 3D (for the same $\taustop$), albeit usually with slightly weaker maximum grain clustering amplitude.

Perhaps most important, these simulations are intended to qualitatively illustrate dust dynamics. We therefore do not attempt to follow cosmological galaxy and/or star formation, nor their enrichment of the medium and/or dust formation, but simply trace a pre-existing grain population in a turbulent disk.

\subsection{Initial conditions}

Without loss of generality, we adopt code units where $\langle c_{s} \rangle=1$, $\langle \rhogas \rangle=1$, and $\Omega=1$ (at the box center). The physics of our problem are then completely specified by three dimensionless parameters: the Mach number $\mathcal{M}$ of the (driven) turbulence, the dimensionless average stopping time, $\langle \taustop \rangle \equiv \rhointernal\,\Rgrain\,\Omega / (\langle \rhogas \rangle\,\langle c_{s} \rangle)$, and the size of the box in code units $L /(c_{s}/\Omega)$. We always choose the driving scale of the turbulence to correspond to a narrow range in Fourier-space (factor $\sim2$ in $k$) centered on the disk scale height $H = \sigma/\Omega = (c_{s}/\Omega)\,(1+\mathcal{M}^{2})^{1/2}$. We adopt an isothermal equation of state (see above), and $q=1$, appropriate for a galactic disk with $V_{c}=$\,constant. Both gas and dust are initialized with uniform density.

Motivated by cosmological simulations of these first galaxies, we focus in this paper on transsonic turbulence with one-dimensional Mach numbers $\mathcal{M}\sim1-2$. And motivated by observed physical grain sizes, we consider $\taustop\sim 0.01,\,0.1,\,1$. Finally, we perform all simulations at $1024^{2}$ resolution. We have considered numerical resolution studies (from $64^{2}-2048^{2}$), and for the $\taustop$ which we simulate here, $1024^{2}$ gives well-converged results (for smaller $\taustop$, however, grains will cluster on still smaller scales, so higher resolution is necessary). We consider two box sizes, with side-length $L=1$ in code units (i.e.\ $L\approx H/\sqrt{1+\mathcal{M}^{2}}$), and $L=5$ ($L\approx 5\,H/\sqrt{1+\mathcal{M}^{2}}$). The former ``zooms in'' on scales we expect to contain $\sim1$ massive GMC complex, while the latter contains many such complexes (but less well-resolved). Because we see good convergence, we analyze the larger boxes in the text since the gas statistics are better sampled, but the results are similar in either case.

We seed the simulations with trace initial magnetic fields; these are amplified by the turbulent dynamo until they saturate around equipartition (magnetic energy about $\sim 5\%$ of the kinetic energy, in good agreement with other simulations using a variety of different numerical techniques and analytic estimates; see \citealt{brandenburg:nonlinear.astro.dynamos,schekochihin:smallscale.turb.dynamo,federrath:supersonic.turb.dynamo}). However because they only indirectly influence the dynamics by (weakly) changing the structure of turbulence, we find that magnetic fields do not significantly change our conclusions compared to hydro-only runs. 

All simulations are run for a timescale at least $\sim20\,\Omega^{-1}$. We discard the first five dynamical times as the statistical properties might be biased by our initial conditions. After about $t\sim \Omega^{-1}$, the galaxy reaches steady-state, and we see no systematic variation in the Mach numbers, magnetic field strength, or distribution of dust/gas densities; we therefore simply average the statistical results over the retained time snapshots for each simulation. Finally, we analyze the results as in \citet{hopkins.lee}, using a local kernel density estimator to measure the distribution of dust and gas densities at all points in the simulation.

\end{appendix}

\end{document}